\documentclass[nofootinbib,preprintnumbers,twocolumn,amsmath,amssymb,showpacs]{revtex4}
\pdfoutput=1
\usepackage{graphicx}
\usepackage{slashed}
\usepackage{footmisc,multirow}
\usepackage{subfigure,enumerate}

\newcommand{\be}{\begin{eqnarray*}}
\newcommand{\ee}{\end{eqnarray*}}

\newcommand{\bee}{\begin{eqnarray}}
\newcommand{\eee}{\end{eqnarray}}
\newcommand{\beeq}{\begin{equation}}
\newcommand{\eeeq}{\end{equation}}

\begin{document}

\title{Dark Sector spectroscopy at the ILC}

\preprint{IPPP/13/61}
\preprint{DCPT/13/122}
\begin{abstract}
Recent studies have shown that searches in the mono-photon and missing energy
final state can be used to discover dark matter candidates at the ILC.
While an excess in this final state over the Standard Model background would indicate the existence of a dark sector, no detailed information
about the internal structure of this sector can be inferred. Here, we demonstrate how just a few observables can discriminate between various realizations of dark sectors, including e.g. the spin of mediators. 
\end{abstract}

\author{Jeppe R.~Andersen} \email{jeppe.andersen@durham.ac.uk}
\affiliation{Institute for Particle Physics Phenomenology , Department
of Physics, Durham University, United Kingdom}
\author{Michael Rauch} \email{michael.rauch@kit.edu}
\affiliation{Institute for Theoretical Physics, University of Karlsruhe,
 Germany}
\author{Michael Spannowsky} \email{michael.spannowsky@durham.ac.uk}
\affiliation{Institute for Particle Physics Phenomenology , Department
of Physics, Durham University, United Kingdom}


\maketitle

\section{Introduction}
\label{sec:intro}
Astronomical observations strongly indicate the existence of dark matter
\cite{Komatsu:2010fb}. Many extensions of the Standard-Model take this
into account by incorporating a so-called dark sector: a sector
of particles that do not carry electric or colour charge. The
interactions of the dark sector can be protected by global symmetries and the
particles can 
have a long lifetime. Often the dark sector is not completely decoupled,
but interacts with the Standard Model particles by the exchange of a Z boson,
or a mediator of a yet unknown force.

In recent years several experiments, i.e. PAMELA \cite{Picozza:2006nm}, Fermi
LAT \cite{FermiLAT:2011ab} and most recently AMS \cite{Aguilar:2013qda}, have
observed an excess in the positron fraction in the electron-positron energy
spectrum at energies above $\sim10$ GeV. A possible explanation for this
excess could be the decay of an invisible particle into  an $e^+ e^-$ pair. Leptophilic dark sectors have been identified as a possible explanation of the observed excess \cite{Fox:2008kb, Cirelli:2008pk, Chen:2008dh, Cohen:2009fz}.

Unfortunately, due to the large uncertainties in evaluating the cosmological backgrounds, indirect detection experiments face challenges in claiming the discovery of a potential dark matter candidate.  
Direct detection experiments try to measure the momentum transfer between the weakly interacting massive particle (WIMP) and the detector. If the WIMP is light, the sensitivity of the experiments are strongly reduced \cite{Savage:2010tg}. Interpretations of searches for WIMPs at direct detection experiments usually assume that the dark sector is minimal, i.e. consists of only one particle, and the WIMP accounts for the total dark matter abundance in the universe. We will not make those strong assumptions. Indeed, if the dark sector is not minimal or the WIMP is not even stable on cosmological time scales constraints from direct detection experiments are strongly relaxed.

Recently it has been pointed out that in case the WIMP couples with reasonable strength to quarks, mono-jet searches at the Tevatron and LHC can be a superior way in discovering them \cite{Birkedal:2004xn,Feng:2005gj, Beltran:2008xg, Cao:2009uw, Beltran:2010ww, Goodman:2010yf, Bai:2010hh, Fox:2011fx, Rajaraman:2011wf}. If the WIMP couples predominantly to leptons constraints can be derived from LEP in mono-photon searches \cite{Fox:2011pm} or a future electron positron collider \cite{Dreiner:2012xm}.

Unfortunately both, direct detection experiments and mono-jet/photon searches, are very limited tools in unravelling the detailed structure of the dark sector, e.g. the spin of the force mediator in combination with its mass and the mass of the WIMP(s). In mono-photon searches the WIMPs recoil against a high pT photon. Therefore, only the total amount of missing transverse energy in the event can be measured, i.e. the differential cross section for one observable.

The observables discussed in this article allow an unbiased view into the dark sector, as long as a mediator couples the dark sector to electrons. Because of the lack of an existing electron-positron collider we will discuss these observables in the context of the International Linear Collider (ILC). So far, the ILC's great potential in studying the structure of dark sectors has not been completely appreciated \cite{ILCTDR}. We assume that the particles in the dark sector are stable on collider time scales and escape detection at the LHC. Therefore, our signal will consist of electrons and missing transverse energy (MET). This signature is relatively rare in the Standard Model, predominantly generated in the production of Z bosons and photons with subsequent decay/splitting to an $e^+e^-$ pair and neutrinos.
An important topology for the study of the structure of the dark sector is the so-called vector boson fusion (VBF) topology (i.e., two possibly forward well-separated electrons), even though the dark sector particles are not necessarily produced by exchanging a vector boson. Tight cuts on the
electron-positron system can reduce the Standard Model backgrounds and increase the average energy flowing through the WIMPs-$e^+e^-$ coupling. A similar strategy is used when studying the coupling structure of the Higgs boson to quarks: Producing a Higgs boson in the VBF channel, angular correlations of the jets can be used to distinguish a CP-even from a CP-odd Higgs boson
\cite{Plehn:2001nj,Klamke:2007cu,Andersen:2010zx} and an invisibly decaying Higgs boson can be disentangled from the backgrounds \cite{Eboli:2000ze}. Similar kinematic configurations can be used to study dark sectors with WIMP-quark couplings at the LHC. However, we find that at the LHC the WIMP-quark coupling has to be of the order of the electroweak coupling to give a significant event shape contribution, reflected in $m_{jj}$ or $\Delta y_{jj}$. Further, at the LHC large systematic uncertainties in final states with missing transverse energy ($\mathrm{MET}$)  and jets render a dark sector spectroscopy a difficult task \cite{An:2013xka}.

This article is organized as follows: In section \ref{sec:eff} we discuss our benchmark models and assumptions which specify the WIMPs-electron-positron interactions. Kinematic observables are identified and interpreted in the context of Regge theory in section \ref{sec:obs}. We further evaluate how well the different benchmark models can be discriminated. In \ref{sec:conc} we present our conclusions.

\section{Benchmark Models and Experimental Constraints}
\label{sec:eff}

For simplicity we assume that the WIMP is a Dirac fermion. The mediator
can be either a scalar particle or a vector particle, which couples only
to electrons. Extending this coupling to all leptons would be
straightforward. The only place where this matters is the width of the
mediator particle, which gets increased by additional couplings to muons
and taus. The width is calculated using the program BRIDGE~\cite{Meade:2007js}.
As the width turns out to be small for the coupling
values considered in the following, taking for example a
generation-blind scenario instead would have no relevant effect on our
results.

The interaction terms appearing in the Lagrangian are denoted in
Table~\ref{tab:operators}.
\begin{table}
\begin{tabular}{l|c|c}
       & scalar & vector \\\hline
e      & $i \ g_{ee\phi,S} \ \bar{e} e \ \phi_S$ 
       & $i \ g_{ee\phi,V} \ \bar{e} \gamma_\mu e \ \phi_V^\mu$ \\
$\chi$ & $i \ g_{\chi\chi\phi,S} \ \bar{\chi} \chi \ \phi_S$ 
       & $i \ g_{\chi\chi\phi,V} \ \bar{\chi} \gamma_\mu \chi \ \phi_V^\mu$ 
\end{tabular}
\caption{Terms in the Lagrangian describing interactions between the
mediator and the electron or the WIMP particle.}
\label{tab:operators}
\end{table}
Our analysis uses light mediator particles. Therefore, the momentum
dependence in the propagator of the mediator plays an important role and
cannot be neglected. Hence, it is not possible to formulate the results
in terms of an effective dimension-six operator of the form
$\bar{e}e \bar{\chi}\chi$ or $\bar{e}\gamma_\mu e \bar{\chi}\gamma^\mu
\chi$ for scalar or vector mediator, respectively, where the mediator is
integrated out.
Nevertheless, one can still define an effective mass $M_*$ as 
\begin{equation}
M_* = \frac{M_\phi}{\sqrt{g_{ee\phi} g_{\chi\chi\phi}}}
\end{equation}
as in Ref.~\cite{Fox:2011fx}.
In the limit of a heavy mediator, the term $1/M_*^2$ becomes the
prefactor of the dimension-six operator.

In the following, we define 8 different model scenarios. They are
characterized by three different options, namely the spin and mass of
the mediator particle and the mass of the WIMP. For the spin of the
mediator particle we investigate the two possibilities already
mentioned, namely a scalar and a vector particle. In many scenarios the dark sector is linked to the Standard Model via kinetic mixing of a dark photon of a hidden $U(1)$ with the $U(1)$ hypercharge of the Standard-Model \cite{Abel:2008ai}. In the scalar case we assume that the mediator couples chiral and exclusively to electrons and WIMPs. By measuring the spin of the mediator such models can be either confirmed or disfavoured. 

 For the two masses we define a light and a heavy scenario each, with values of 8 GeV and 200
GeV for the mediator particle and 5 and 120 GeV for the WIMP mass.

An overview is shown in Table~\ref{tab:models} together with the
effective mass $M_*$ used for each scenario.
\begin{table}
\begin{tabular}{l||r|r|r|r}
model & mediator mass & mediator spin & WIMP mass & 
 \multicolumn{1}{|c}{$M_*$} \\\hline
LSL   &         8 GeV & 0 (scalar)    &     5 GeV &    30 GeV \\
LVL   &         8 GeV & 1 (vector)    &     5 GeV &    30 GeV \\
LSH   &         8 GeV & 0 (scalar)    &   120 GeV &  27.4 GeV \\
LVH   &         8 GeV & 1 (vector)    &   120 GeV &    21 GeV \\
HSL   &       200 GeV & 0 (scalar)    &     5 GeV &  1250 GeV \\
HVL   &       200 GeV & 1 (vector)    &     5 GeV &  1250 GeV \\
HSH   &       200 GeV & 0 (scalar)    &   120 GeV & 332.4 GeV \\
HVH   &       200 GeV & 1 (vector)    &   120 GeV & 511.8 GeV \\
\end{tabular}
\caption{Overview of the different model scenarios used in our analysis.
The first and third letter of the model name denote the mass (light or
heavy) of the mediator and WIMP, respectively, while the middle one
describes the spin nature of the mediator (scalar or vector).}
\label{tab:models}
\end{table}
The exact choice of masses is somewhat arbitrary, but has been guided by
the following considerations. The mass of the light WIMP is chosen such
that it is below the typical reach of direct detection experiments,
while the heavy scenario has a mass which is beyond the reach of direct
searches at LEP. The two choices for the mediator mass have then been
chosen such that in the light-light and heavy-heavy models the on-shell
decay of a mediator particle into two WIMPS is kinematically forbidden.

The coupling parameters for the light WIMP models are already
constrained by direct searches at LEP. Therefore, we choose our
effective mass such that they are at the 90\% CL exclusion boundary
given in Ref.~\cite{Fox:2011fx}. For the 200 GeV mediator the effective
mass parameter can be taken directly from there, while for the 8 GeV
mediator we have instead used the given value for the 10 GeV curve. This
is slightly more restrictive than the true 8 GeV value, so with this
choice we are erring on the safe side. 
The heavy WIMP scenarios with a WIMP mass of 120 GeV are beyond the
reach of LEP. As the mediator couples only to electrons, also searches
at hadron colliders cannot significantly constrain the coupling
parameters. Only a direct interpretation of the WIMP as dark matter
candidate would immediately yield strong constraints by direct detection
experiments~\cite{Savage:2010tg}, and in fact reduce the signal-to-background
ratio to a value too low for realistic studies. Therefore, we set the
couplings in the heavy WIMP cases to a value that gives similar cross
sections as the corresponding light WIMP scenario. This choice also
simplifies comparisons between the two options.

\section{Discriminative observables}
\label{sec:obs}
\begin{figure}[htbp]
  \centering
  \includegraphics[width=\columnwidth]{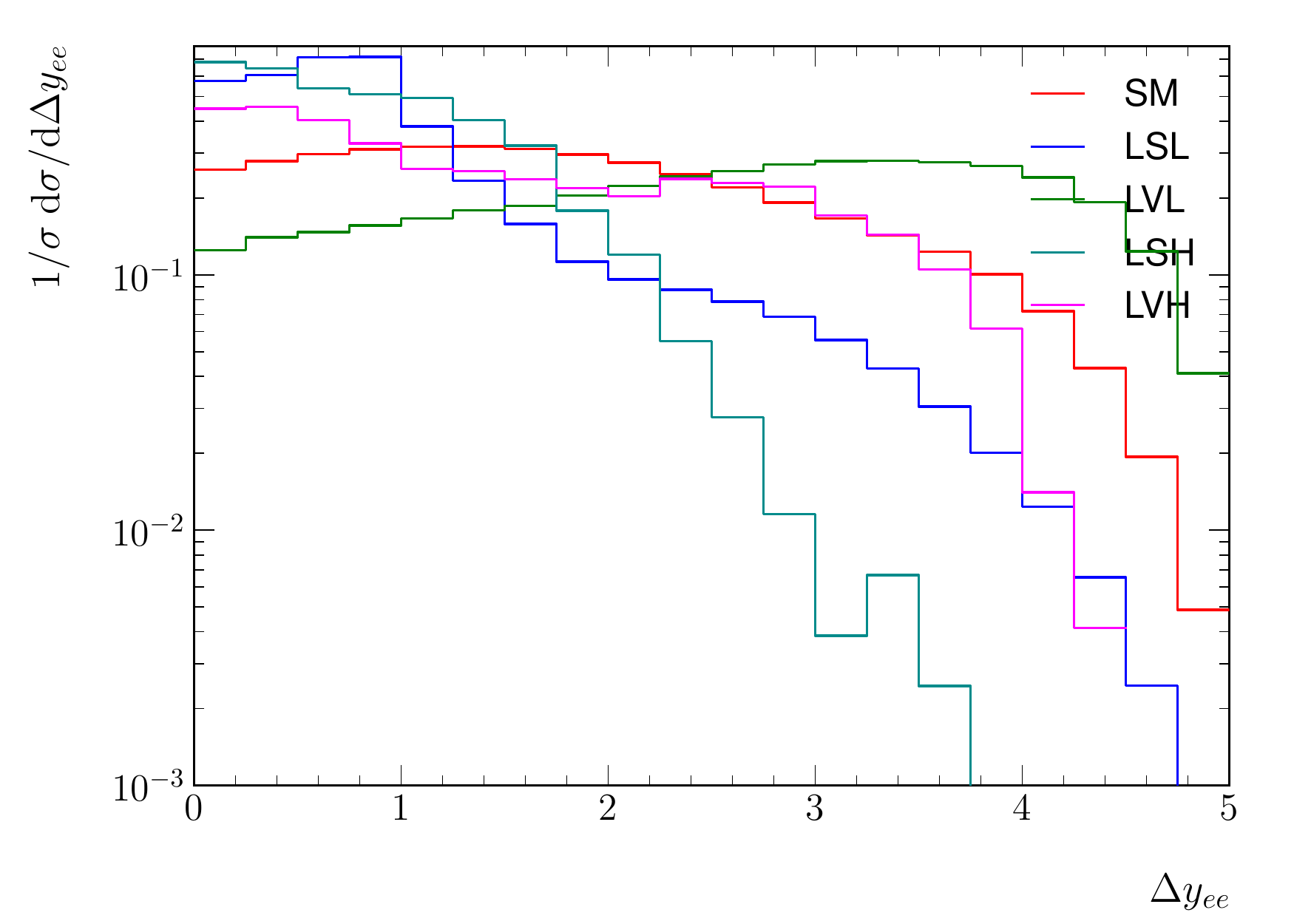}
  \includegraphics[width=\columnwidth]{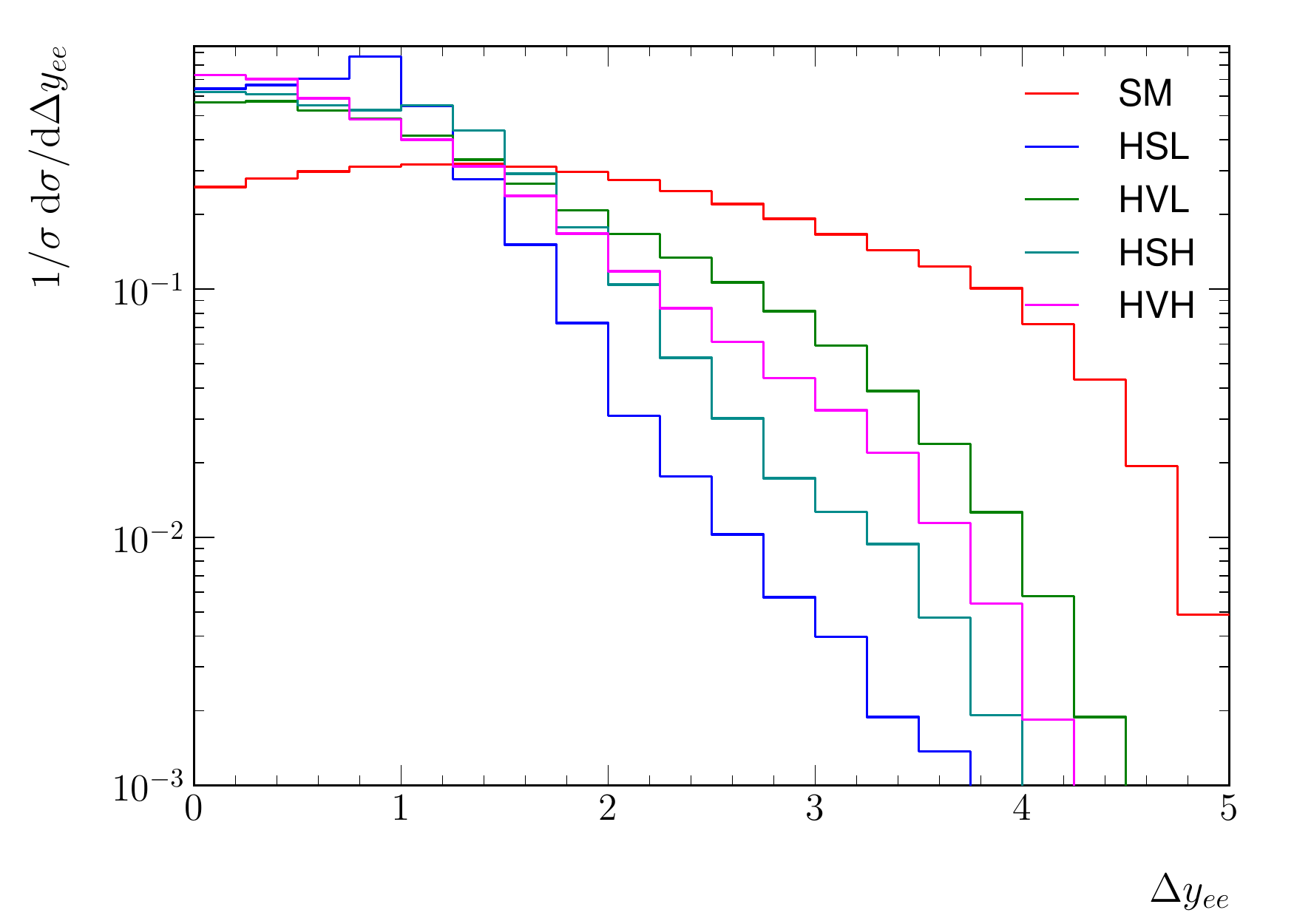}
  \caption{$1/\sigma \mathrm{d}\sigma/\mathrm{d}\Delta y_{ee}$ for the
    standard model, and the models with light or heavy dark matter candidates
    and (top) a light mediator, (bottom) a heavy mediator. See text for more
    details.}
  \label{fig:dsigmaddeltay} 
\end{figure}

The spin of the mediator can be assessed by appealing to the analytic
behaviour of the scattering amplitude dictated by Regge theory~\cite{Regge:1959mz,Brower:1974yv} in the
limit of large invariant mass between each produced particle compared to the
propagating momentum (and any mass of fields), $s_{ij}\gg|t_i|$. In this
\emph{multi-Regge kinematic} limit, which is attained within the VBF cuts,
the analytic behaviour of a $2\to n$ scattering is determined by
\begin{align}
  \mathcal{M}^{p_a p_b\to p_1 p_2 p_3 p_4}\to s_{12}^{\alpha_1(t_1)}\
  s_{23}^{\alpha_2(t_2)}\ s_{34}^{\alpha_3(t_3)}\ \gamma,
\end{align}
$p_1,\ldots,p_4$ are the final state momenta ordered in rapidity, and $\gamma$ depends on the
couplings, the $t$-channel momenta $t_i$ and ratios of $s_{ab}/(\prod
s_{ij})$ only. The powers $\alpha_i$ determining the scaling behaviour with
$s_{ij}$ depend on the spin of the particle exchanged in the $t$-channel,
$\alpha_i=J_i$ up to radiative corrections. In cases where the mass of the
exchanged particles is negligible, the spin of the exchanged particle can
therefore be inferred by studying the scaling of the cross section with the
invariant mass between the electron/positron pair, see
Fig.~\ref{fig:dsigmaddeltay}. Since $s_{ab}=2 p_{a\perp}p_{b\perp}
\left(\cosh(y_a-y_b)-\cos(\phi_a-\phi_b)\right)$, the constraint on the
analytic behaviour of the scattering amplitude means that the spin of the
exchanged messenger particle can be directly probed by investigating the
scaling of the cross section with either the rapidity difference or the
invariant mass between the electron-positron pair.

However, if the exchanged particle has a mass, which is large compared to the
other scales of the process, then there will be modifications to this simple
picture. Other distributions would then be consulted to differentiate the
spin and the mass simultaneously.

\begin{figure}[htbp]
  \centering
  \includegraphics[width=\columnwidth]{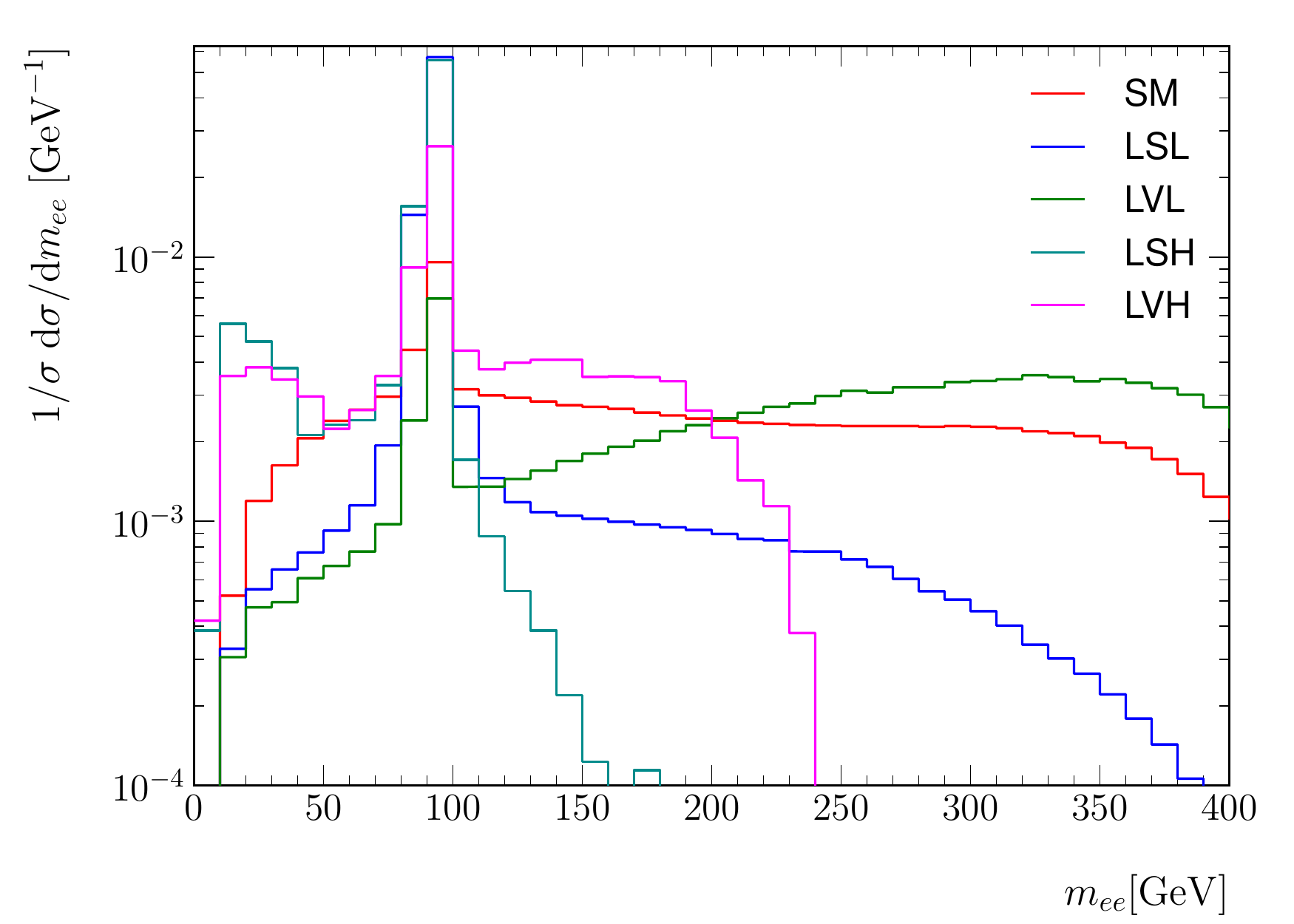}\\
  \includegraphics[width=\columnwidth]{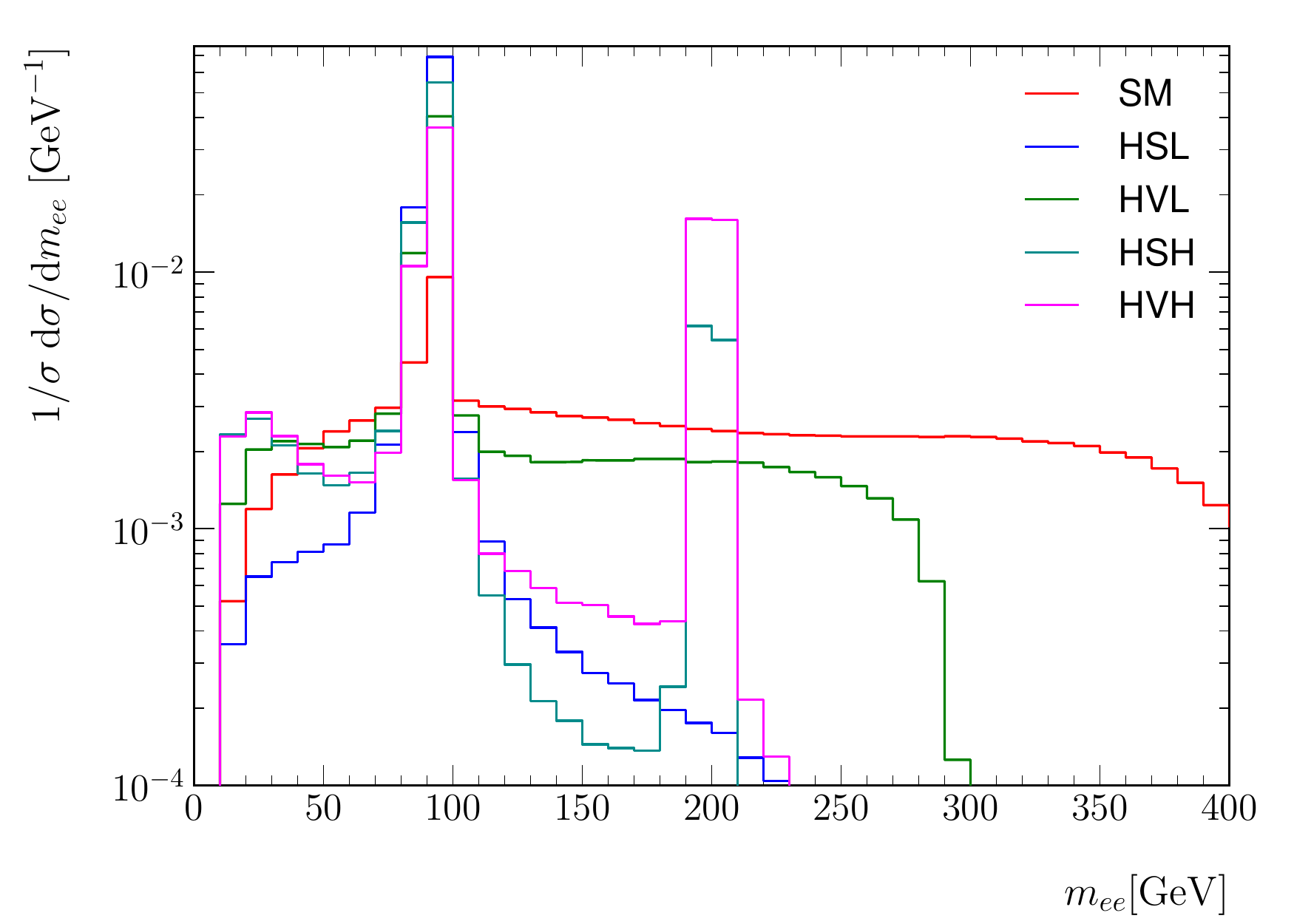}\\
  \caption{$1/\sigma \mathrm{d}\sigma/\mathrm{d}m_{ee}$ for the standard model,
    and the models with light or heavy dark matter candidates and (top) a
    light mediator, (bottom) a heavy mediator. See text for more details.}
  \label{fig:dsigmadmee} 
\end{figure}

Fig.~\ref{fig:dsigmaddeltay} illustrates the different scaling of the cross
section with $m_{ee}$ and $\Delta y_{ee}$ respectively, within our
models. The predictions from the Regge analysis is observed: for a fixed
setup of masses, the scalar exchange is suppressed at large $m_{ee}$ and
$\Delta y_{ee}$ compared to the models with a vector mediator.  While the
impact of the heavy mediator mass is significant for a fixed mass of the dark
matter candidate, the dominance at large $\Delta y_{ee}$ of vector exchanges
over scalar exchanges still holds as predicted by the Regge analysis. The
same is true for $1/\sigma \mathrm{d}\sigma/\mathrm{d}m_{ee}$
(Fig.~\ref{fig:dsigmadmee}). For polarised beams and vector mediators, the
$S/B$ can reach 27\% at large $\Delta y$, which is where the vector like
signal processes will peak.

Conversely, scalar exchange models can get $S/B$ enhanced by studying only
regions of small $\Delta y_{ee}$. 

\begin{figure}[htbp]
  \centering
  \includegraphics[width=\columnwidth]{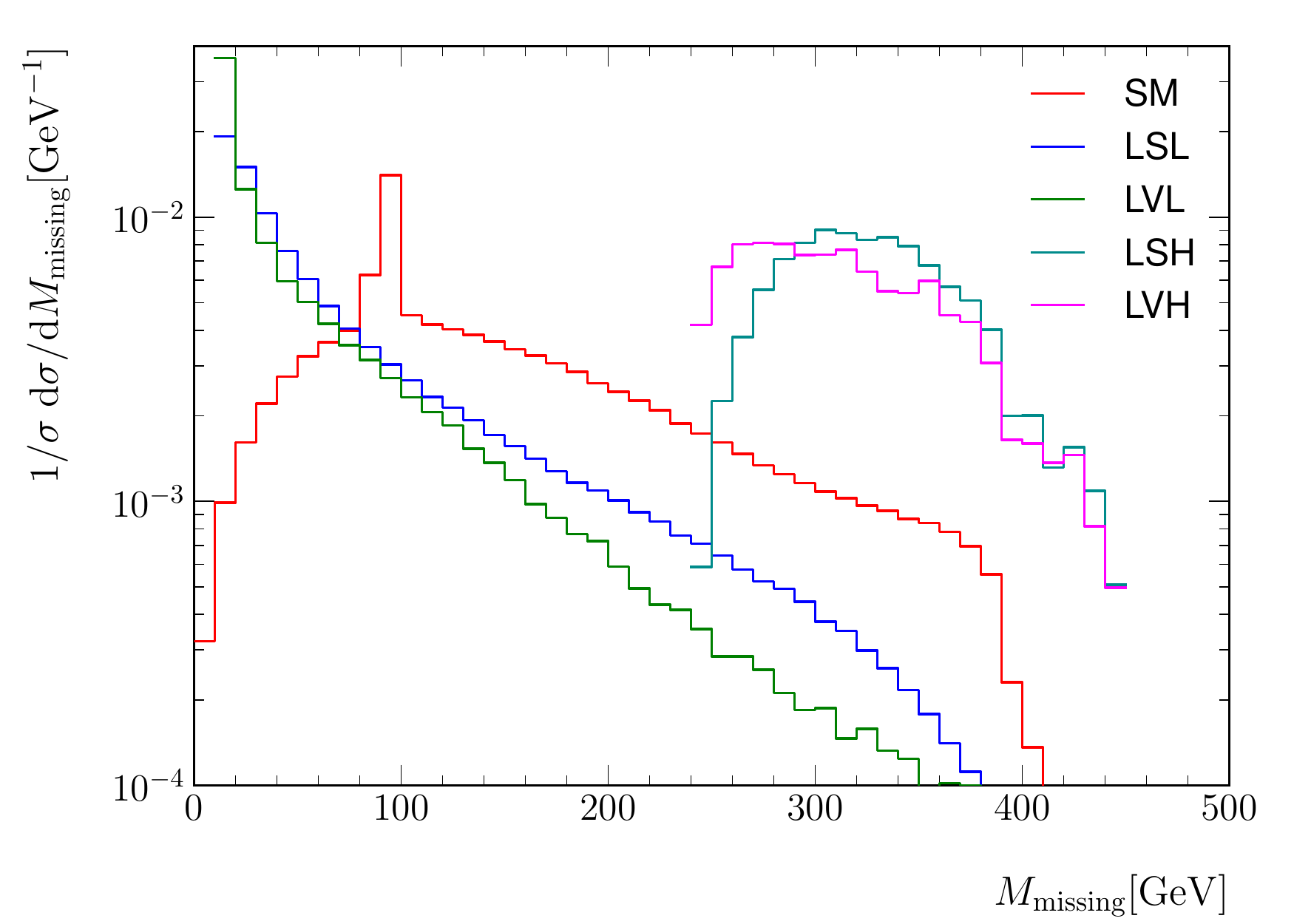}
  \includegraphics[width=\columnwidth]{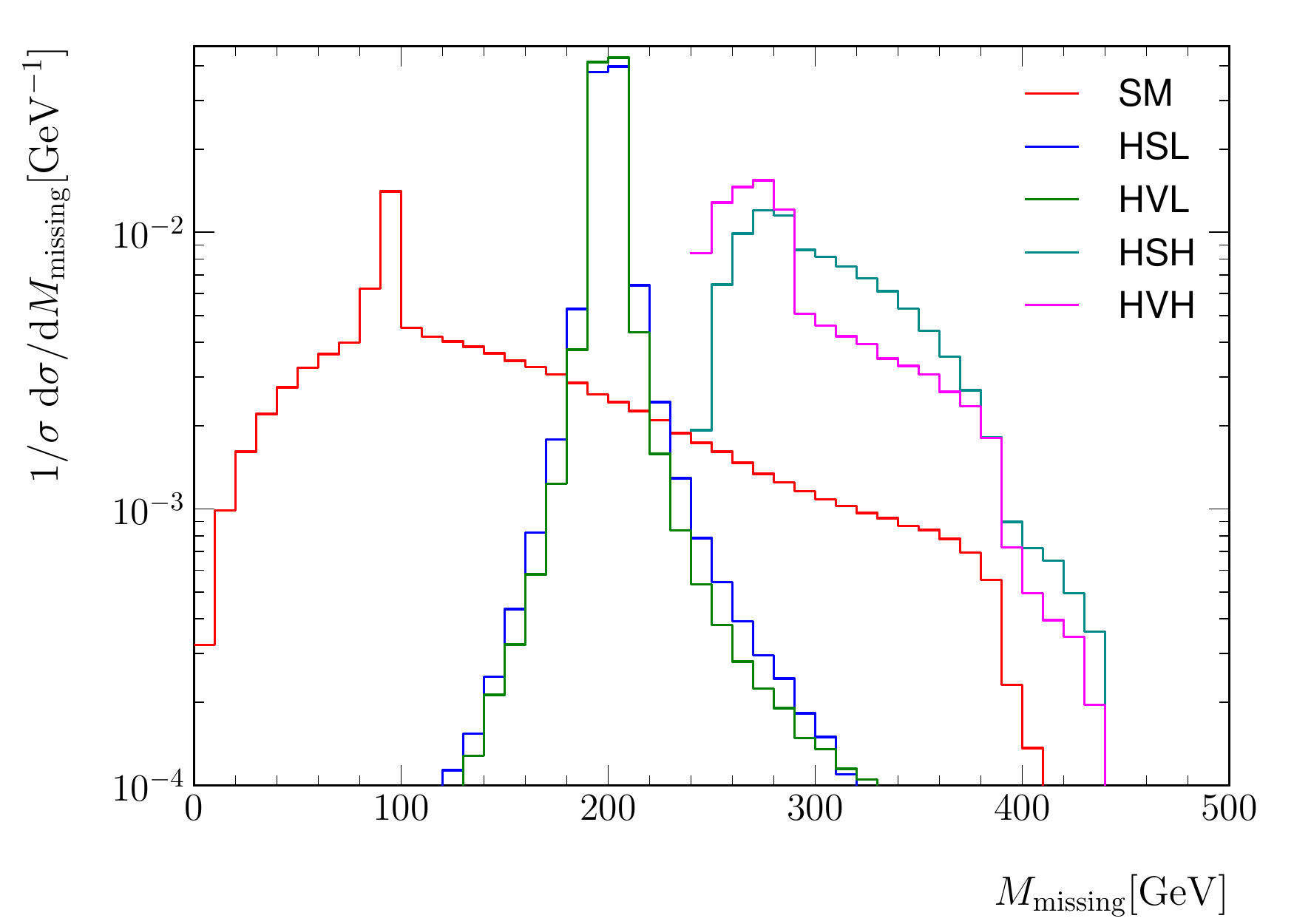}
  \caption{$1/\sigma\ \mathrm{d}\sigma/\mathrm{d}M_{\mathrm{missing}}$ for
    the standard model, and the models with light or heavy dark matter
    candidates and (top) a light mediator, (bottom) a heavy mediator. See
    text for more details.}
  \label{fig:dsigmadmmissing}
\end{figure}

In Fig.~\ref{fig:dsigmadmmissing} we plot the normalised differential
spectrum of the invariant mass of the invisible 4-momentum in the event. This
is obviously bounded from below by the sum of the masses of the two dark
matter particles. If this bound is below the mass of the mediator, then the
spectrum has a pronounced peak at this mass for both scalar and vector
mediators. The shape of the spectrum clearly identifies the mass-hierarchy of
the mediator and DM particles: When the bound from the DM particles is above
the mass of the mediator, the spectrum is very broad, as contrasted with the
pronounced peak at the mediator mass.

In case we studied scalar CP-even or CP-odd WIMPs, $\Delta \phi_{e^-e^+}$ can be helpful for their discrimination. For the $8$ models we study here, $\Delta \phi_{e^-e^+}$ is of minor importance.  

In conclusion, the distribution of the invariant mass of the invisible
momentum, $M_\mathrm{missing}$ can uniquely determine the mass scale and
hierarchies of the mediator and dark matter particle, but does not
discriminate between scalar and vector mediators. However, once the mass
scales are determined, the spin of the mediator can be determined from the
shape of the cross section with respect to $\Delta y_{ee}$ or equivalently
$m_{ee}$, due to the spin-dependence dictated by the Regge-analysis. Other
distributions on e.g.~the transverse momentum of the hardest lepton can then
be used to check for consistency.

\section{Discussion}
\label{sec:conc}

As discussed in Section~\ref{sec:obs} we expect the observables $m_{e^+e^-}$ and the invariant mass of the WIMPS, precisely measured as recoil system of $e^+e^-$, to be the strongest discriminators for the benchmark models. 
At the ILC the lepton's energy is precisely determined and the beams can be partly polarized. We assume conservatively that the degree of polarization is $80 \%$ for the electrons and $30 \%$ for the positrons. In Table~\ref{tab:cs} we show the cross section for 3 different beam polarizations: unpolarized, $++$ and $+-$. The fist (second) index refers to the electron (positron) beam. We find that the ratio between signal and background cross section can be improved for all models by using polarized beams.

\begin{table}[tb]
\begin{tabular}{l||c|c|c}
model & $\sigma_{\mathrm{unpol}}$ & $\sigma_{\mathrm{++}}$ & $\sigma_{\mathrm{+-}}$  \\\hline
SM    &  115.8  &  49.1     &    36.4    \\
LSL   &   1.60  &  1.79    &   1.40  \\
LVL   &   15.07 &  12.80    &  17.02  \\
LSH   &   1.45 &  1.80  &  1.10 \\
LVH   &   9.99 &  7.64  &   12.33  \\
HSL   &   1.17 &  1.43  &   0.92  \\
HVL   &   0.85 &  0.71  &   0.89 \\
HSH   &  1.18 &  1.45  &   0.90 \\
HVH   &  0.85 &  0.64  &   0.98  \\
\end{tabular}
\caption{Cross sections in femtobarn for the different models imposing the constraints as outlined in the text. The three columns refer to the three different beam polarizations: averaged, $++$ and $+-$.  The fist index refers to the electron beam of which we assume to be able to polarize $80\%$ (always $+$). The second index represents the positron beam of which we assume to have a polarization of $30\%$ . The cross sections vary between ($0.7-13.0 \% $, $1.3-26.1 \%$, $2.4-46.8 \% $) of the Standard Model background for three polarizations (all, $++$, $+-$) respectively. }
\label{tab:cs}
\end{table}

Note, the signal models' cross sections respond differently to a change in the polarization of the beams. Hence the inclusive production cross section can be used to discriminate between the models as well, i.e. for scalar mediators $\sigma_{++} > \sigma_{\mathrm{unpol}}$ while for vector mediators $\sigma_{+-} > \sigma_{\mathrm{unpol}}$. However, in this work we will focus on the observables identified in Section~\ref{sec:obs} only and because of the recent interest in vector mediators \cite{Abel:2008ai} we will choose the $+-$ beam polarization in the following. 

\begin{figure*}[!t]
\centering
\subfigure[][~Missing transverse energy]{
\includegraphics[width=0.43\textwidth]{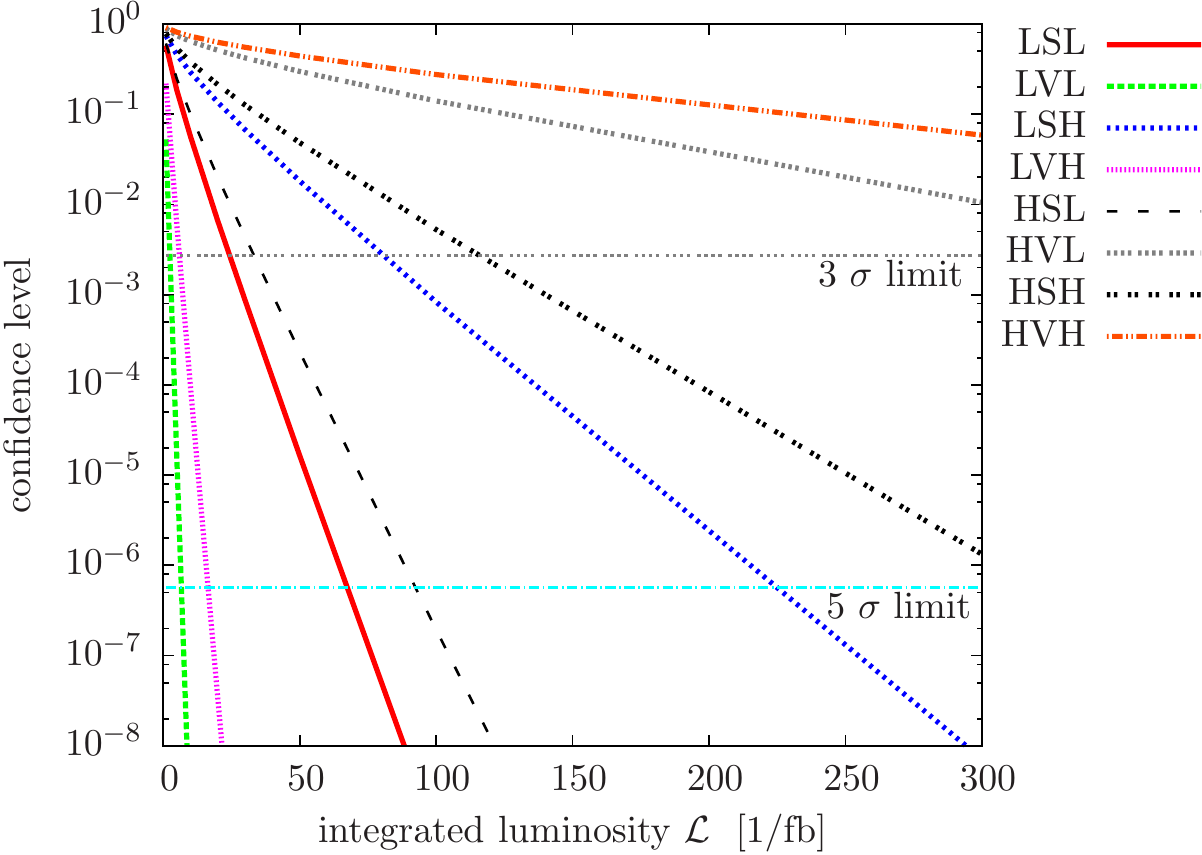} 
}
\hspace{0.2cm}
\subfigure[][~Invariant mass of $e^+e^-$ system]{
\includegraphics[width=0.43\textwidth]{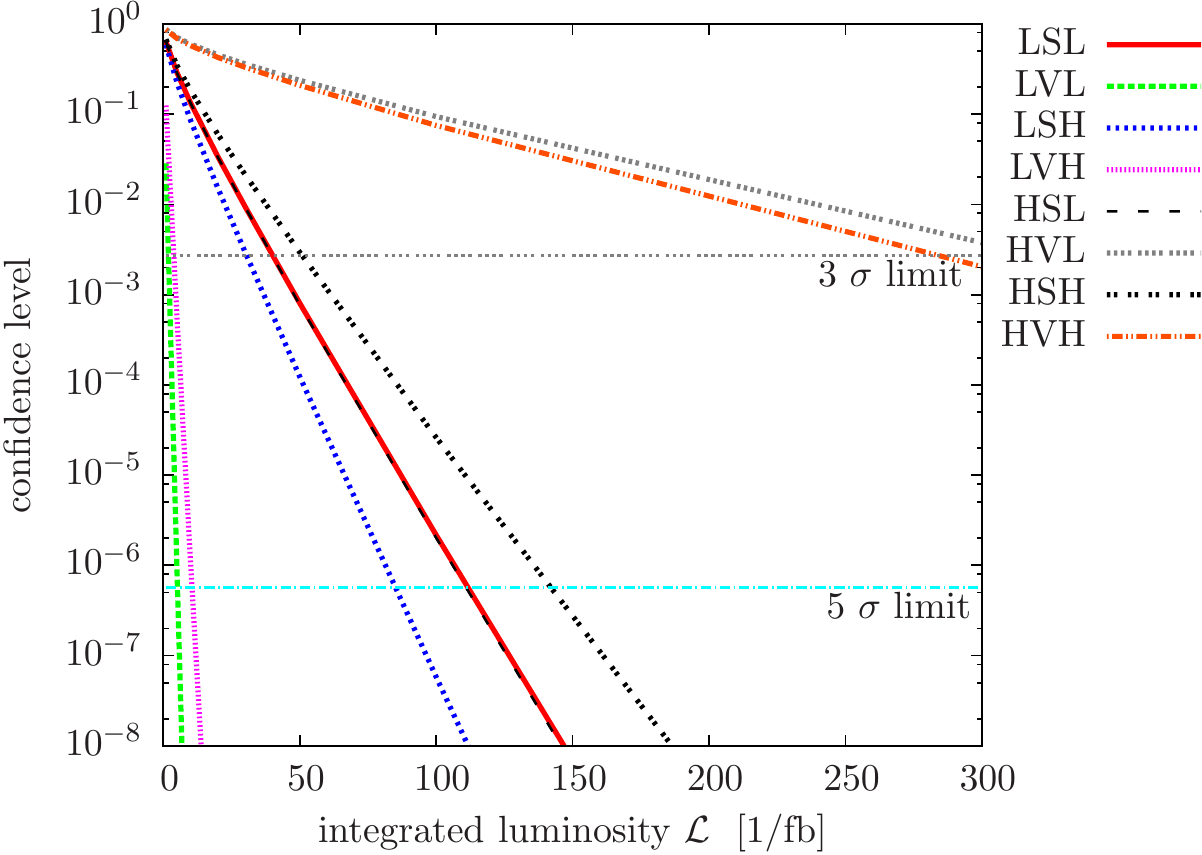}  
} \\
\vspace{0.3cm}
\subfigure[][~$\Delta \phi_{e^+e^-}$]{
\includegraphics[width=0.43\textwidth]{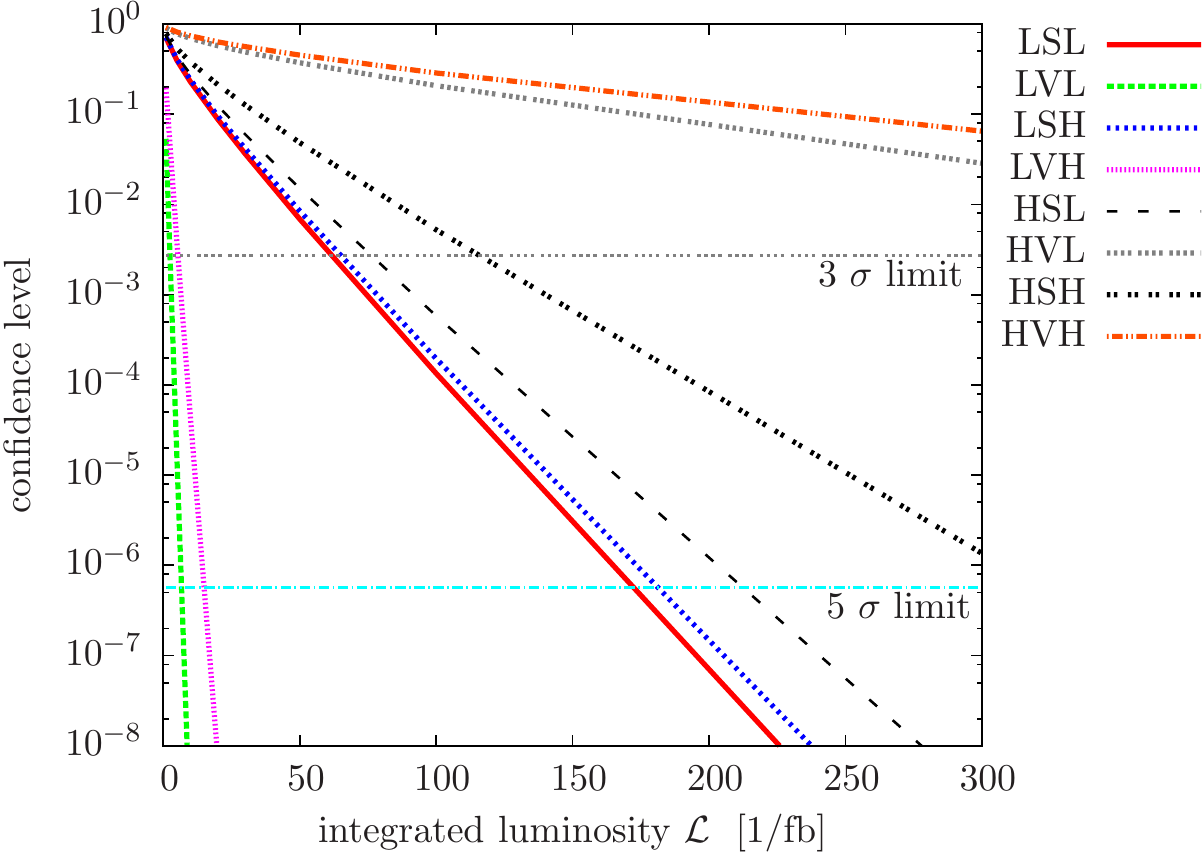} 
}\hspace{0.2cm}
\subfigure[][~$\Delta y_{e^+e^-}$]{
\includegraphics[width=0.43\textwidth]{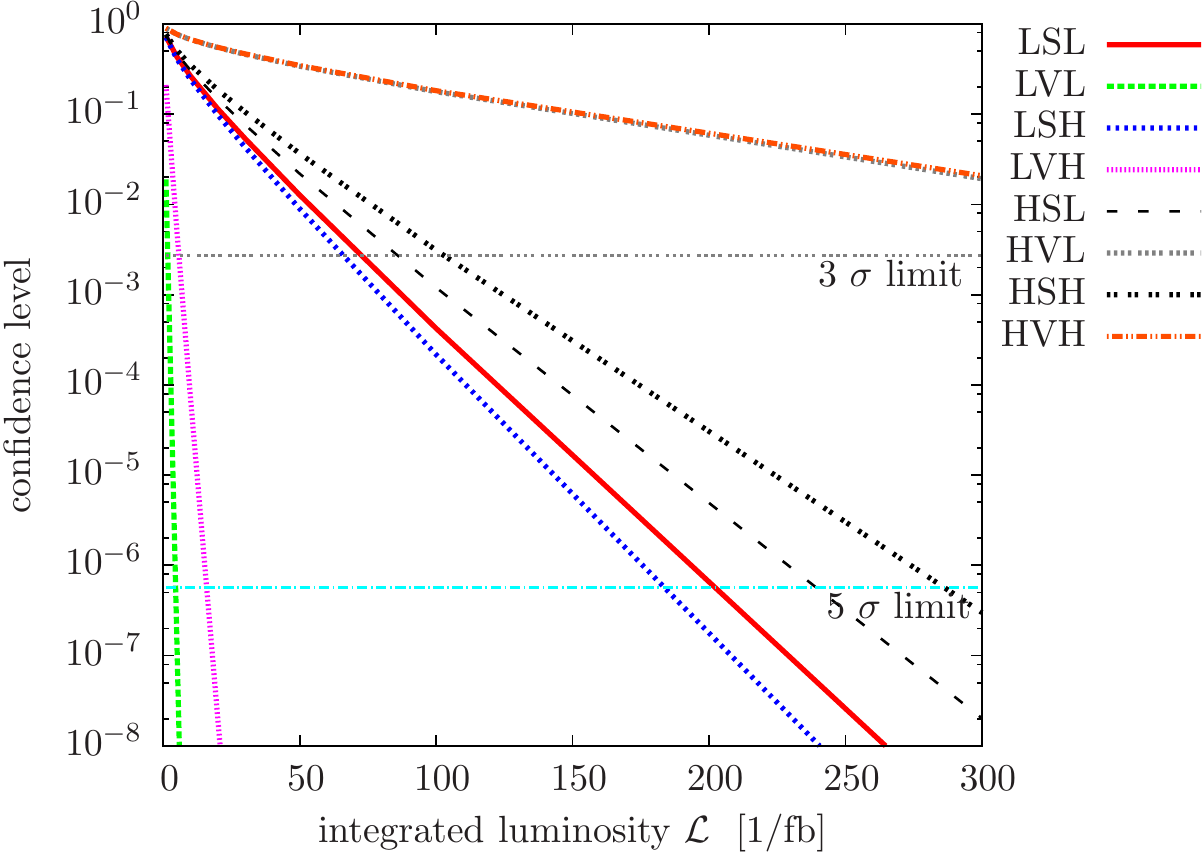} 
} \\
\vspace{0.3cm}
\subfigure[][~$p_T$ of hardest lepton]{
\includegraphics[width=0.43\textwidth]{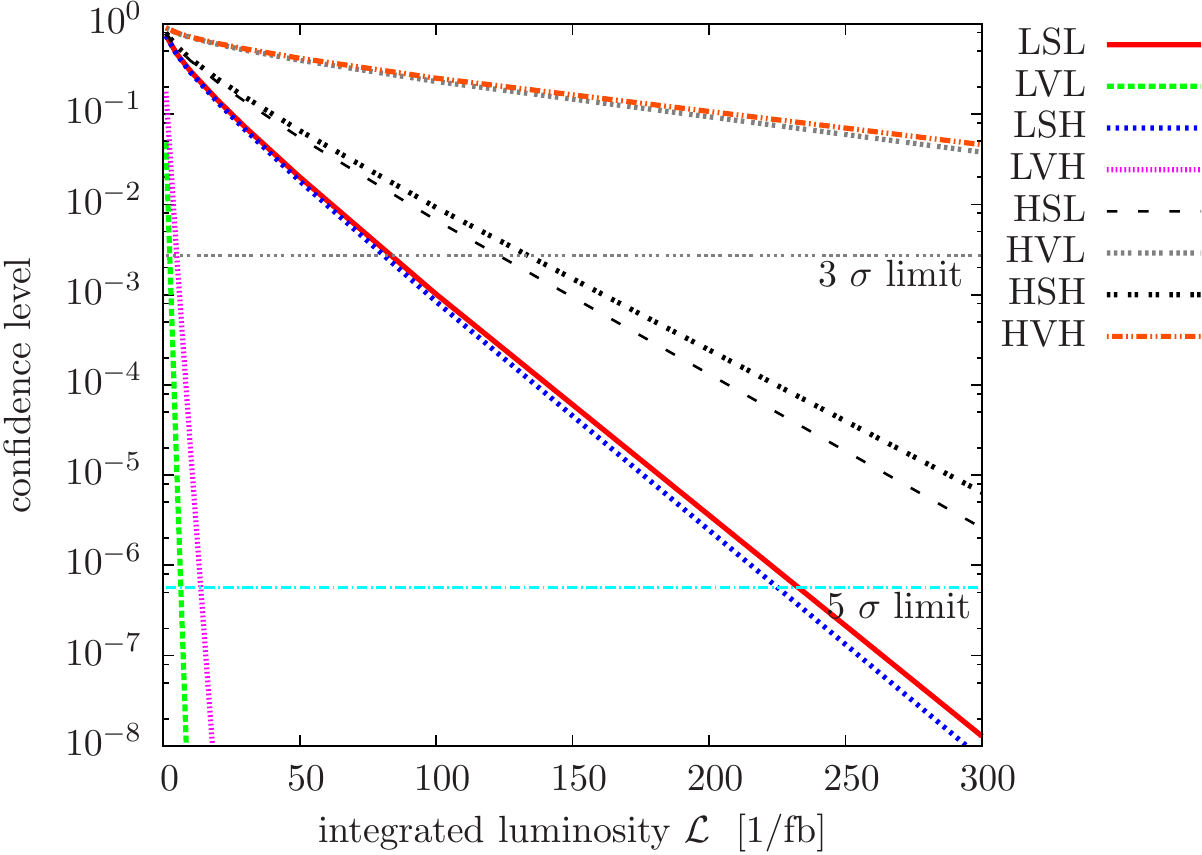}
}\vspace{0.3cm}
\subfigure[][~Invariant mass of WIMP system, $M_{\mathrm{missing}}$]{
\includegraphics[width=0.43\textwidth]{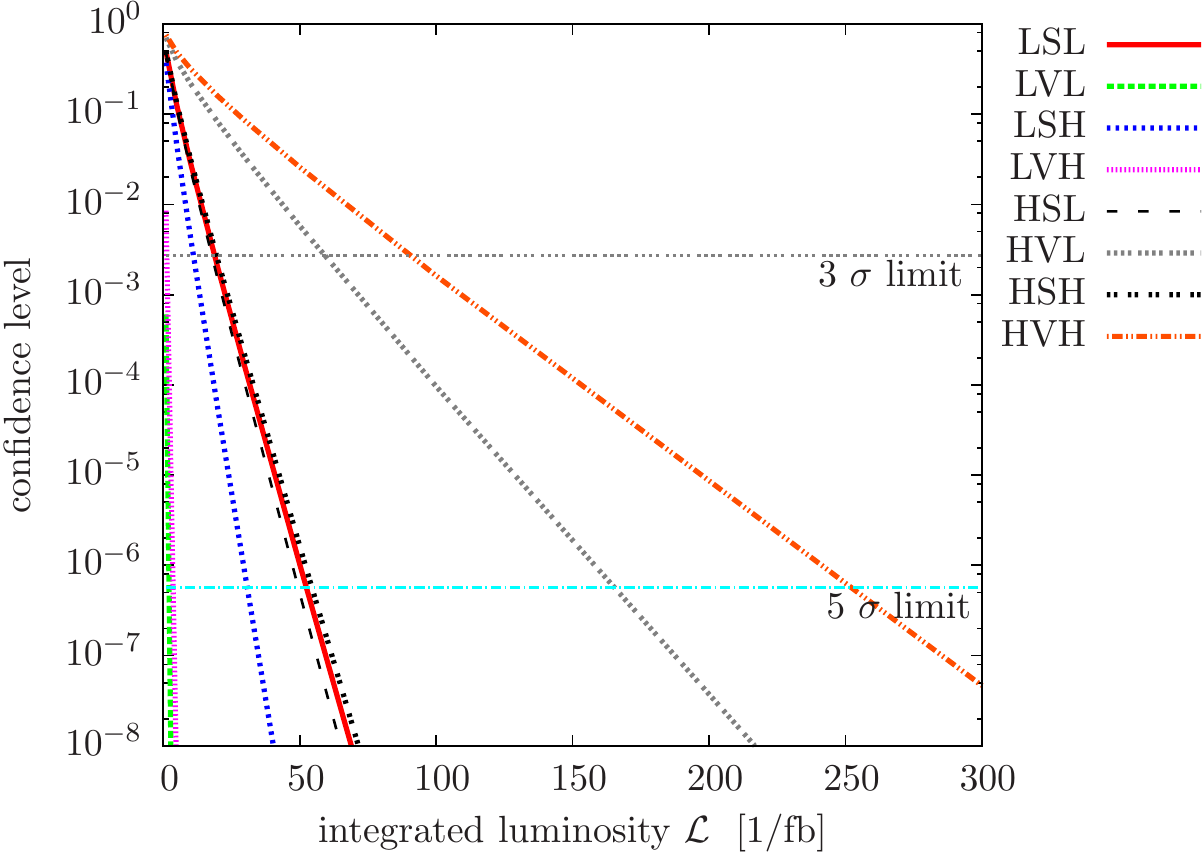}
}
 \caption{Assuming realization of the Standard Model only, these plots show the confidence level to which the respective Dark Sector model can be disfavored using just one observable. The curves take into account not only the shapes but also the normalization of the model's cross section.}
\label{fig:CLSM} 
\end{figure*}

We perform a binned log-likelihood hypothesis test~\cite{llhr} using the CL$_S$ method~\cite{Read:2002hq} to evaluate how well the 8 benchmark models can be discriminated from each other and from the Standard Model. 
For the graphs in Fig.~\ref{fig:CLSM} we assume only the Standard Model is realized in nature. Due to the large allowed cross section of the LVL and LVH models they can be excluded for the given coupling strength with less than $10~\mathrm{fb}^{-1}$. However, with an integrated luminosity of roughly $300~\mathrm{fb}^{-1}$ all models can be disfavored at the $5\sigma$ confidence level.

\begin{figure*}[!t]

\centering
\subfigure[][~$M_{\mathrm{missing}}$]{
\includegraphics[width=0.43\textwidth]{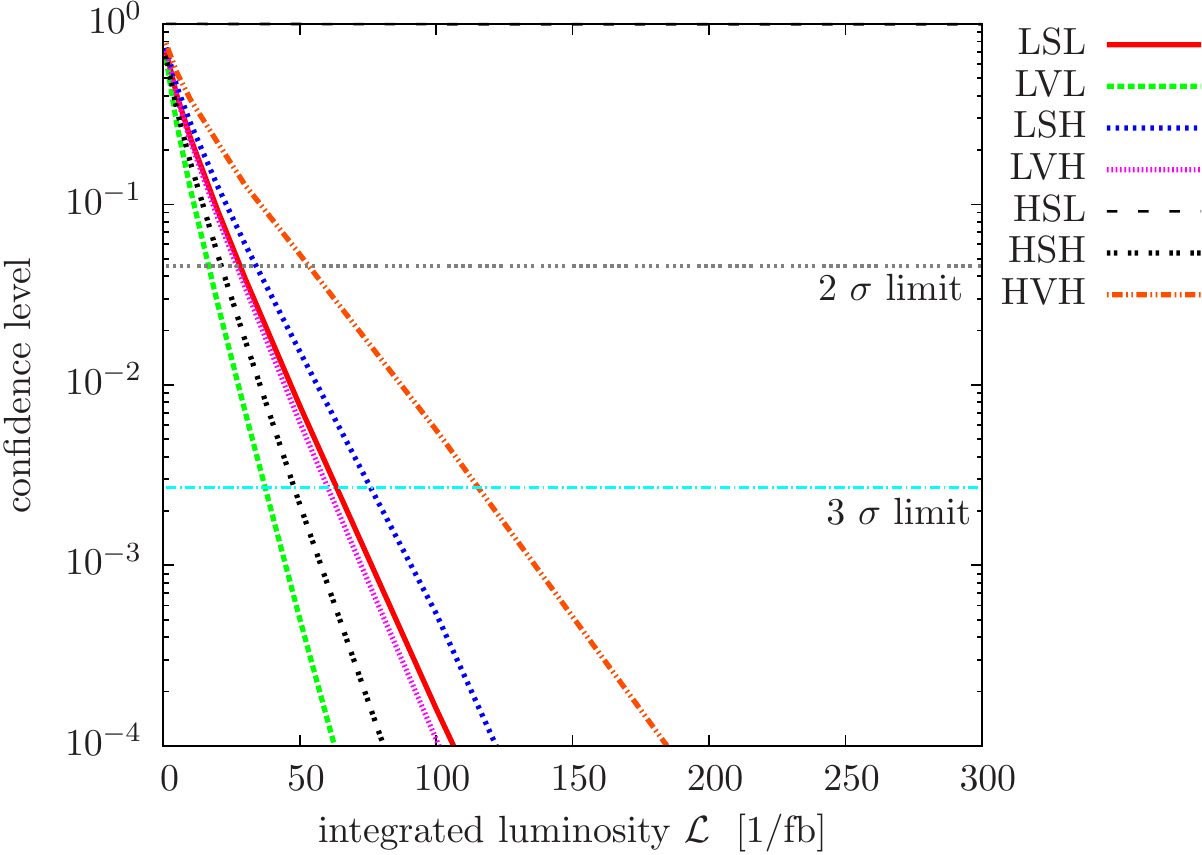} 
\label{sfiga25}
}
\hspace{0.2cm}
\subfigure[][~Invariant mass of $e^+e^-$ system]{
\includegraphics[width=0.43\textwidth]{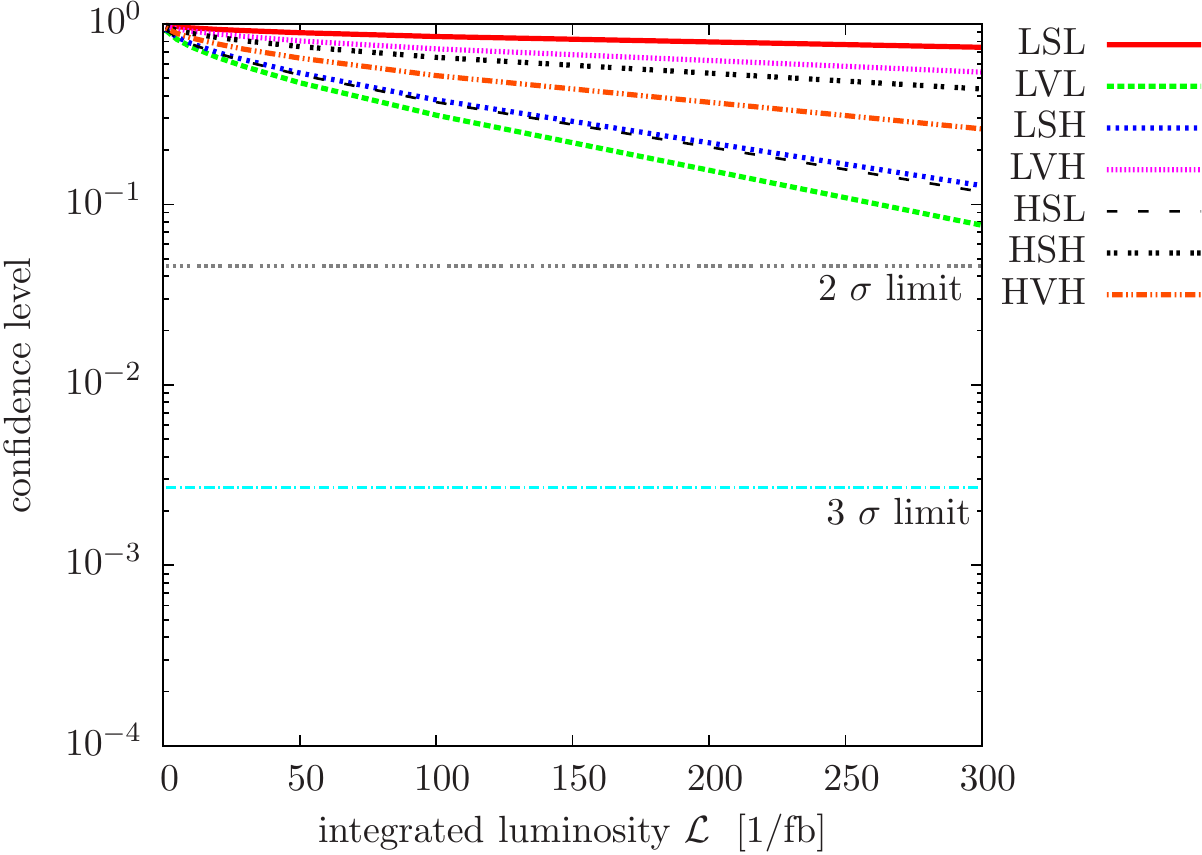} 
\label{sfigb25}
}
\subfigure[][~$M_{\mathrm{missing}}$]{
\includegraphics[width=0.43\textwidth]{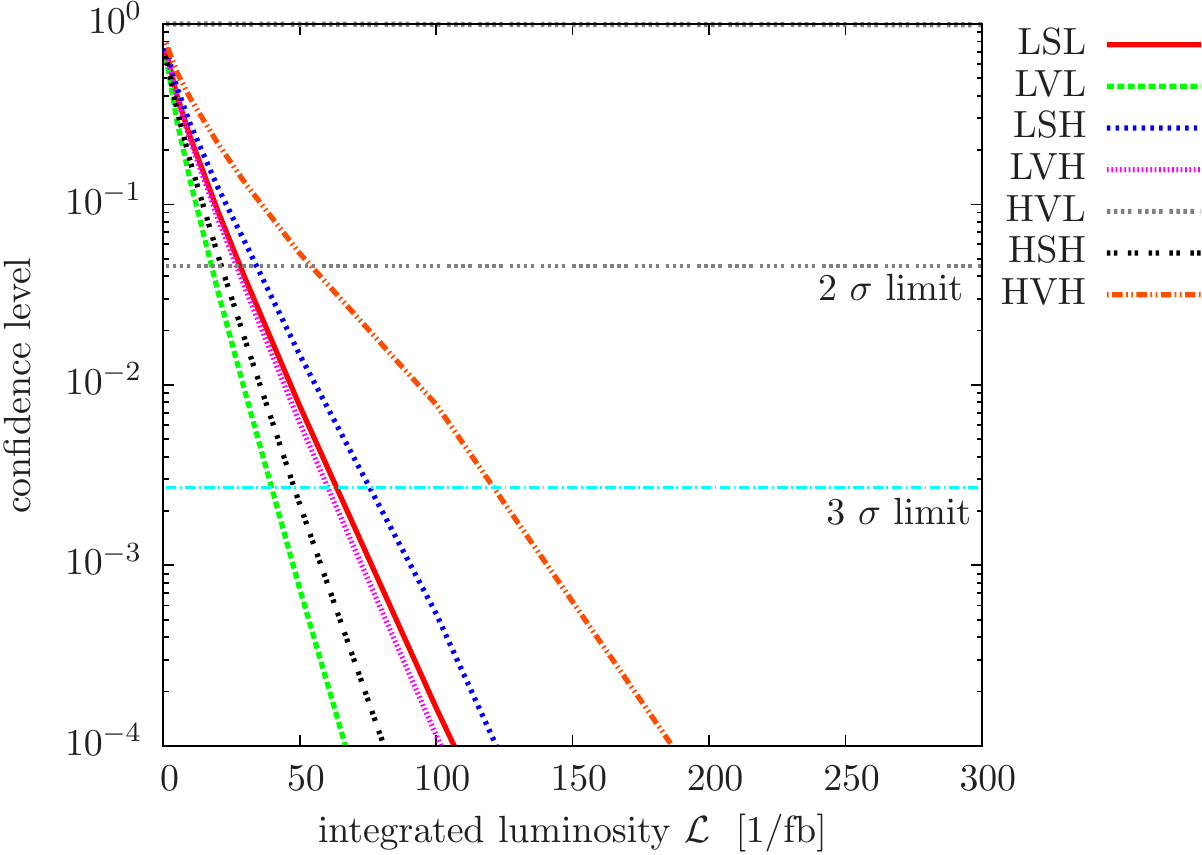} 
\label{sfigc25}
}
\hspace{0.2cm}
\subfigure[][~Invariant mass of $e^+e^-$ system]{
\includegraphics[width=0.43\textwidth]{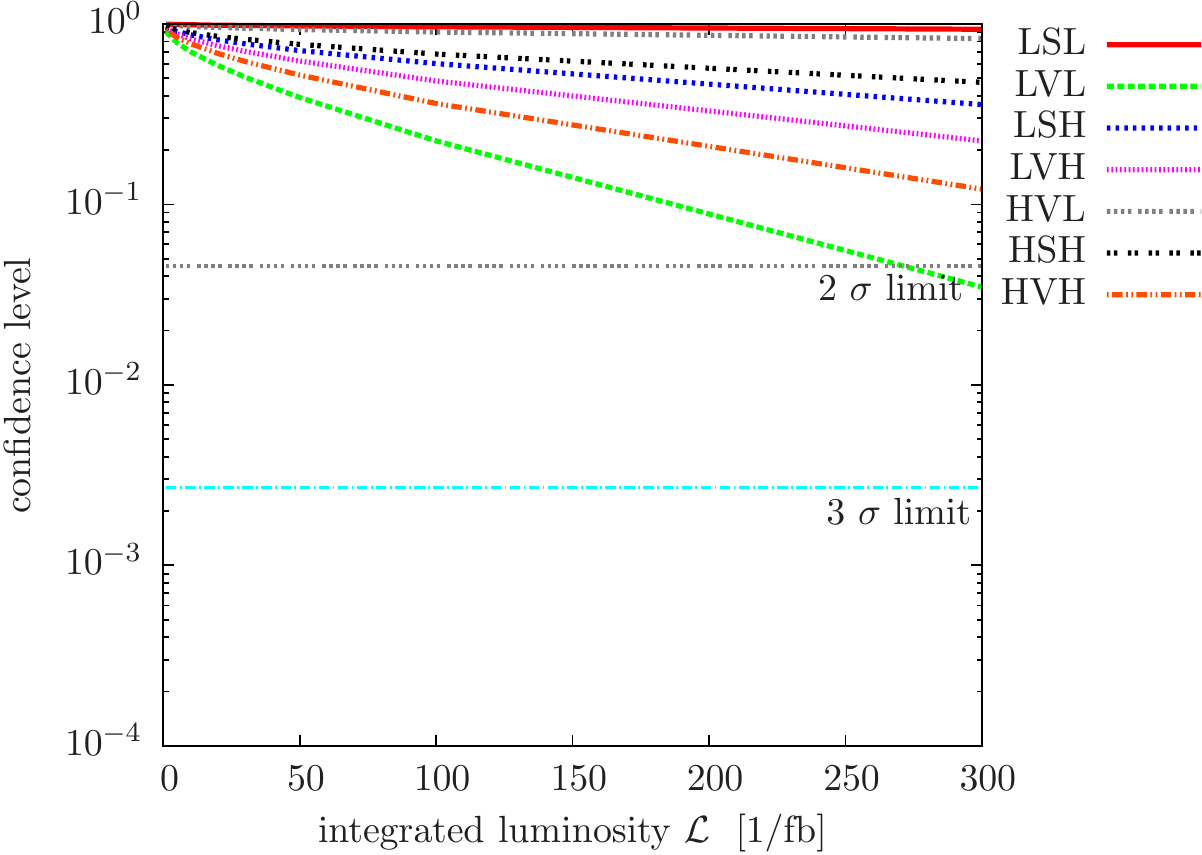} 
\label{sfigd25}
}
\subfigure[][~$M_{\mathrm{missing}}$]{
\includegraphics[width=0.43\textwidth]{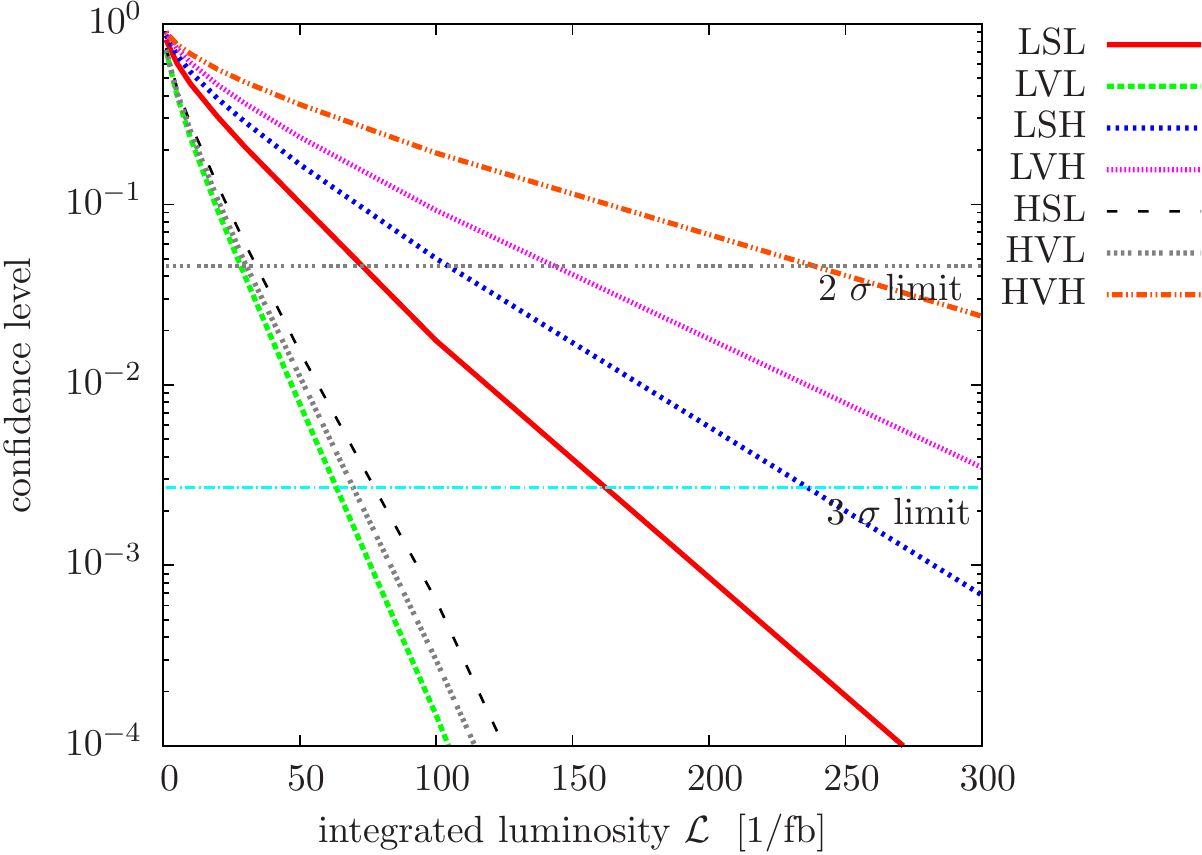} 
\label{sfige25}
}
\hspace{0.2cm}
\subfigure[][~Invariant mass of $e^+e^-$ system]{
\includegraphics[width=0.43\textwidth]{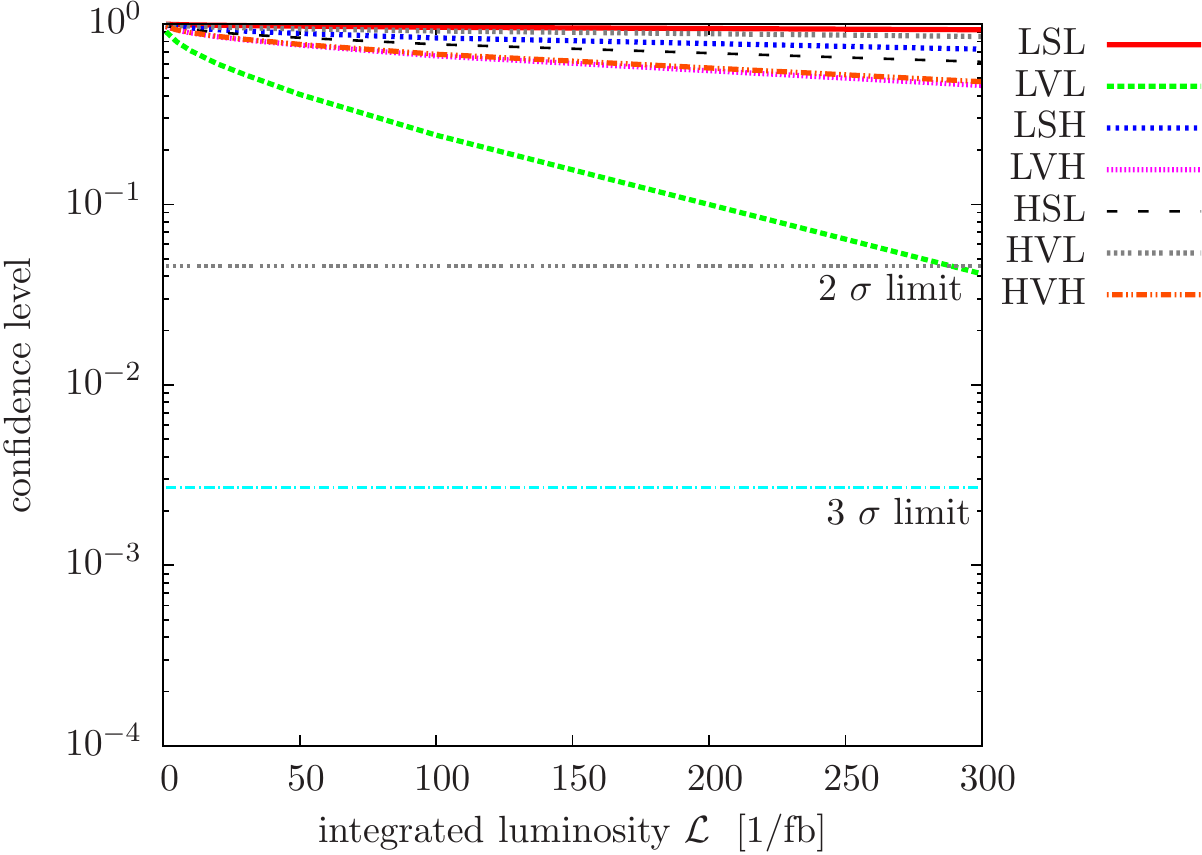} 
\label{sfigf25}
}
 \caption{The curves illustrate the level to which the $M_{\mathrm{missing}}$ and $m_{e^+e^-}$ can discriminate between the different models, assuming the realization of HVL (\ref{sfiga25}-\ref{sfigb25}), HSL (\ref{sfigc25}-\ref{sfigd25}) and HSH (\ref{sfige25}-\ref{sfigf25}). All cross sections are assumed to be $2.5 \%$ of the Standard Model background cross section.}
\label{fig:CLHSH} 
\end{figure*}

\begin{figure*}[!t]
\centering
\subfigure[][~$M_{\mathrm{missing}}$]{
\includegraphics[width=0.43\textwidth]{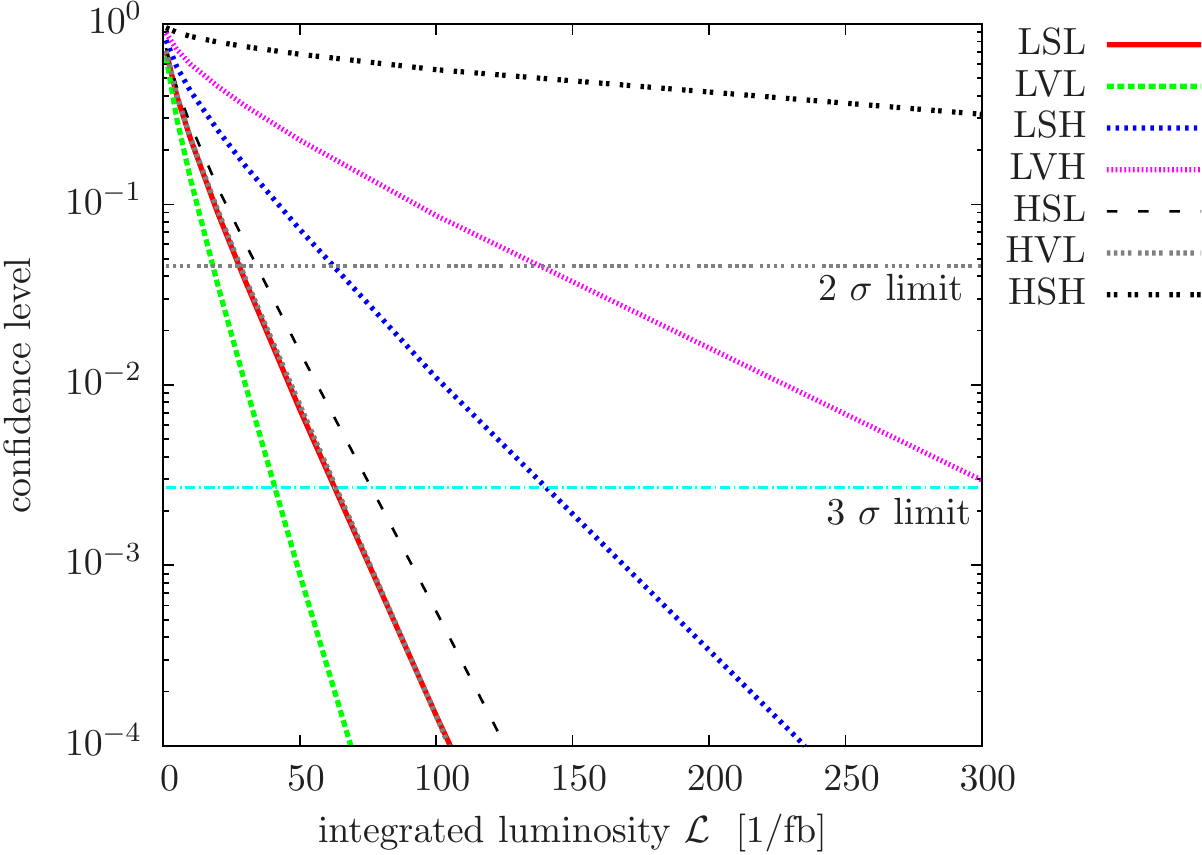} 
\label{sfigg25}
}
\hspace{0.2cm}
\subfigure[][~Invariant mass of $e^+e^-$ system]{
\includegraphics[width=0.43\textwidth]{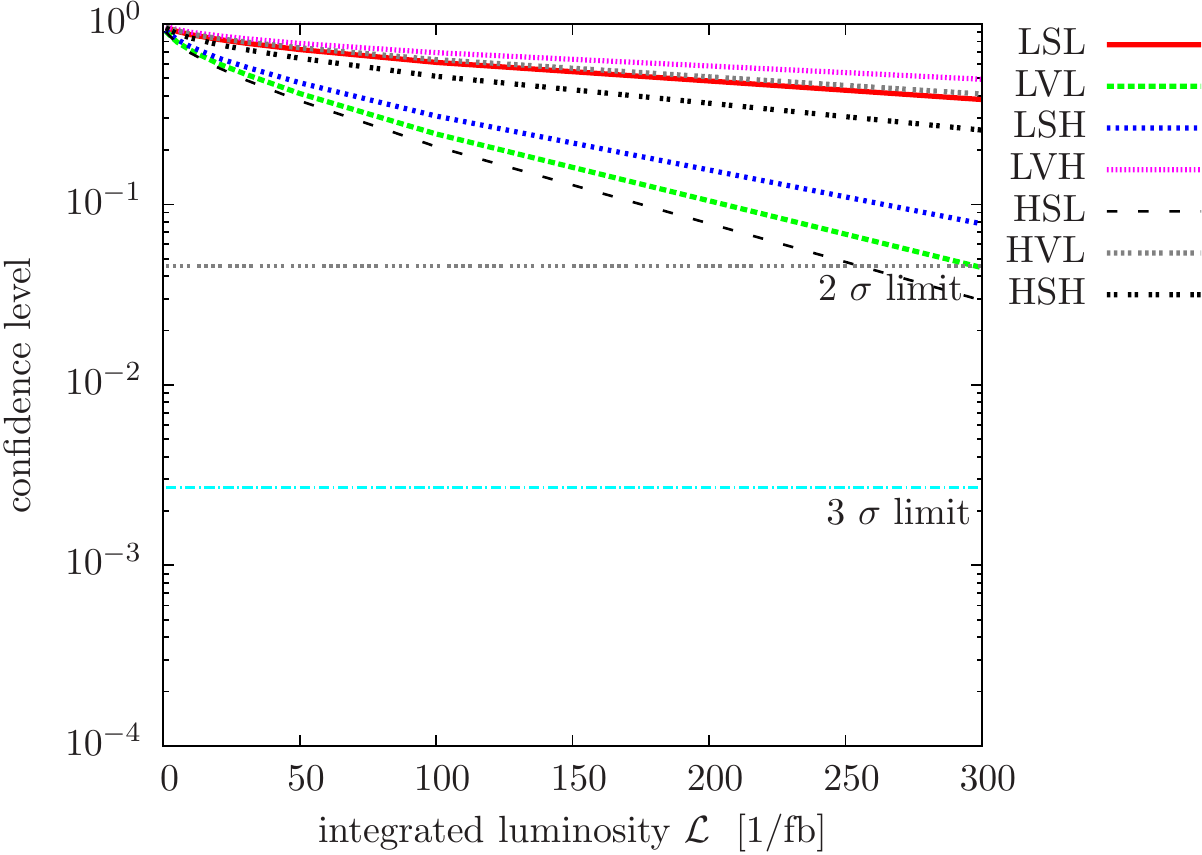} 
\label{sfigh25}
}
\subfigure[][~$M_{\mathrm{missing}}$]{
\includegraphics[width=0.43\textwidth]{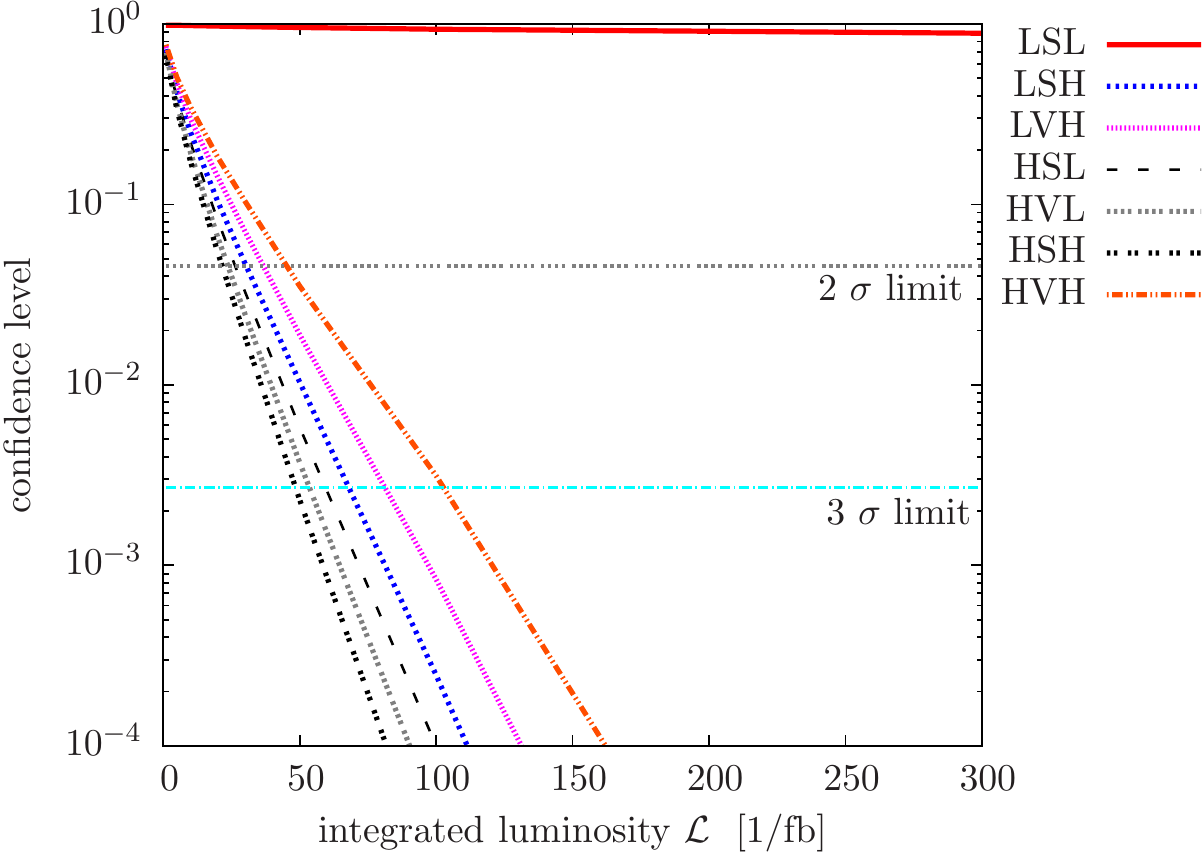} 
\label{sfigi25}
}
\hspace{0.2cm}
\subfigure[][~Invariant mass of $e^+e^-$ system]{
\includegraphics[width=0.43\textwidth]{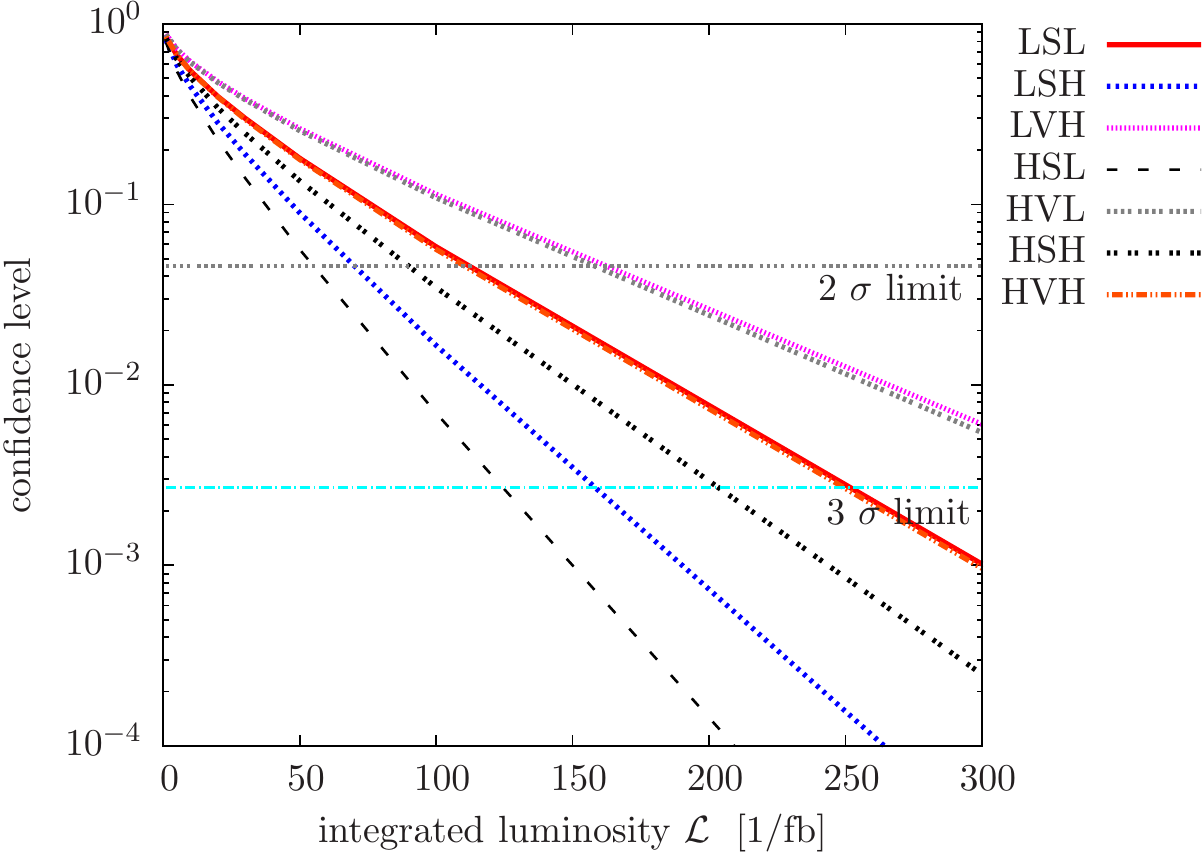} 
\label{sfigj25}
}
\subfigure[][~$M_{\mathrm{missing}}$]{
\includegraphics[width=0.43\textwidth]{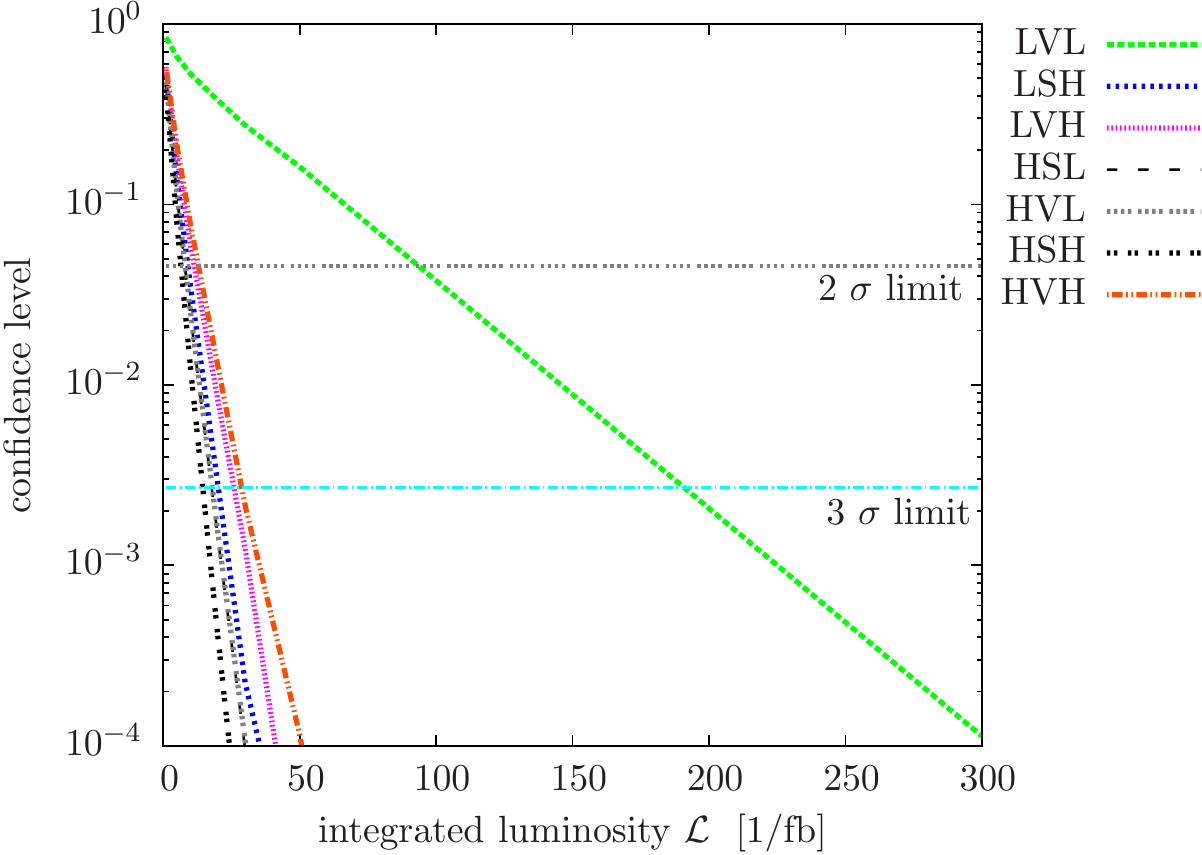} 
\label{sfigk25}
}
\hspace{0.2cm}
\subfigure[][~Invariant mass of $e^+e^-$ system]{
\includegraphics[width=0.43\textwidth]{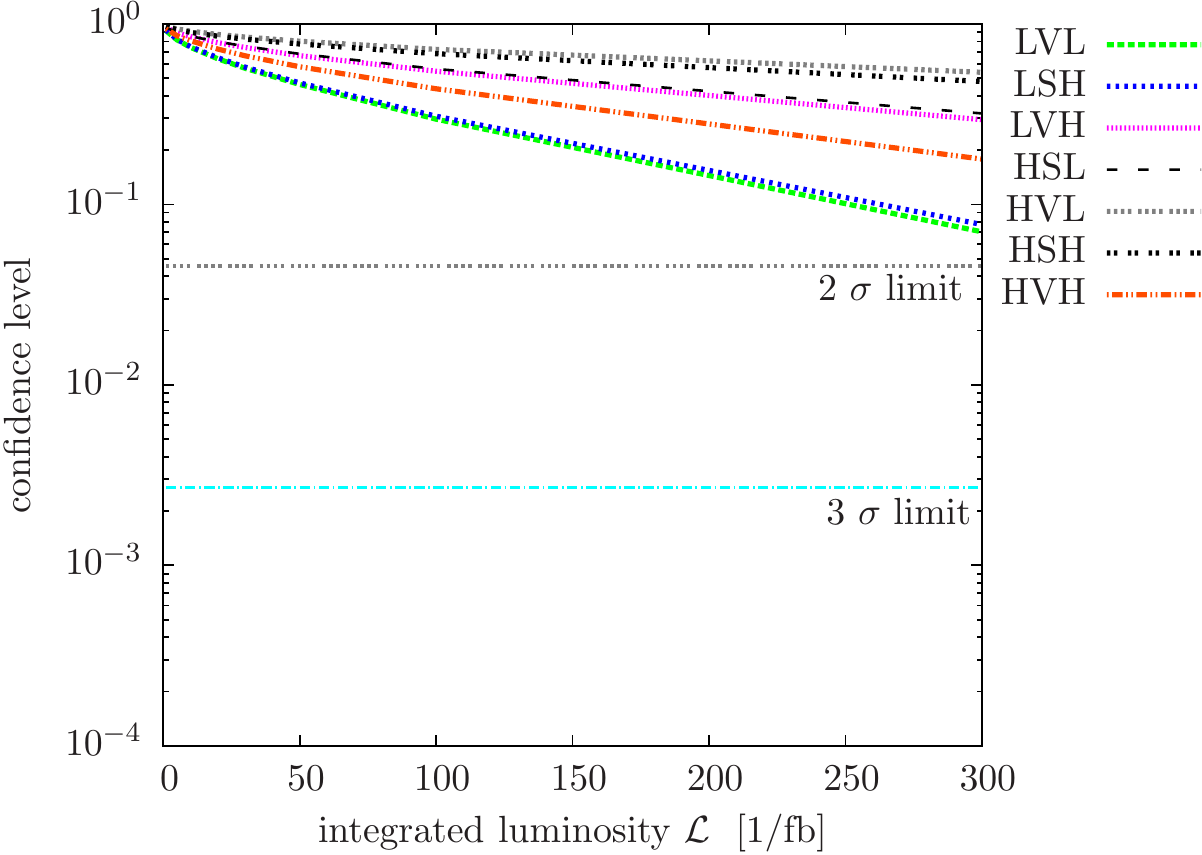} 
\label{sfigl25}
}
 \caption{The curves illustrate the level to which the $M_{\mathrm{missing}}$ and $m_{e^+e^-}$ can discriminate between the different models, assuming the realization of LVL (\ref{sfigg25}-\ref{sfigh25}), LSL (\ref{sfigi25}-\ref{sfigj25}) and LVH (\ref{sfigk25}-\ref{sfigl25}). All cross sections are assumed to be $2.5 \%$ of the Standard Model background cross section.}
\label{fig:CLLSL} 
\end{figure*}

\begin{figure*}[!t]
\subfigure[][~$M_{\mathrm{missing}}$]{
\includegraphics[width=0.43\textwidth]{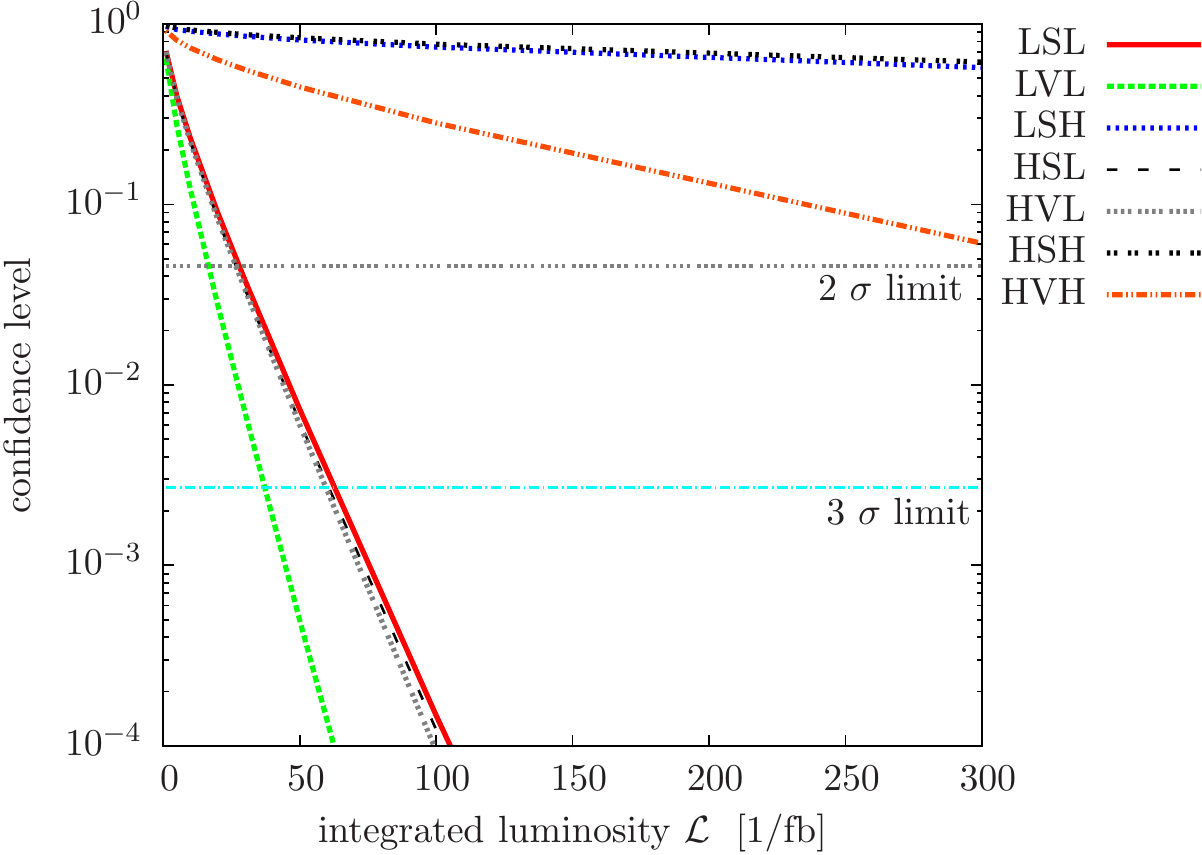} 
\label{sfigm25}
}
\hspace{0.2cm}
\subfigure[][~Invariant mass of $e^+e^-$ system]{
\includegraphics[width=0.43\textwidth]{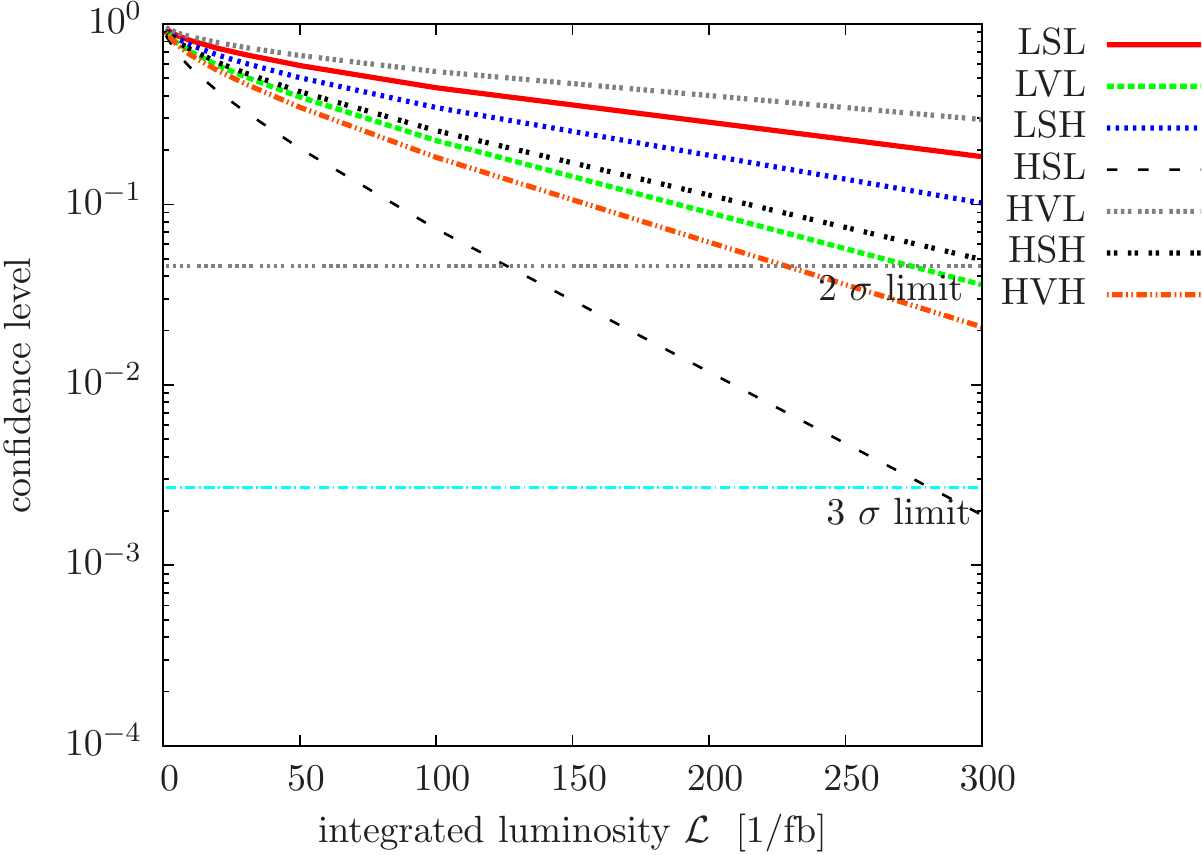} 
\label{sfign25}
}
\subfigure[][~$M_{\mathrm{missing}}$]{
\includegraphics[width=0.43\textwidth]{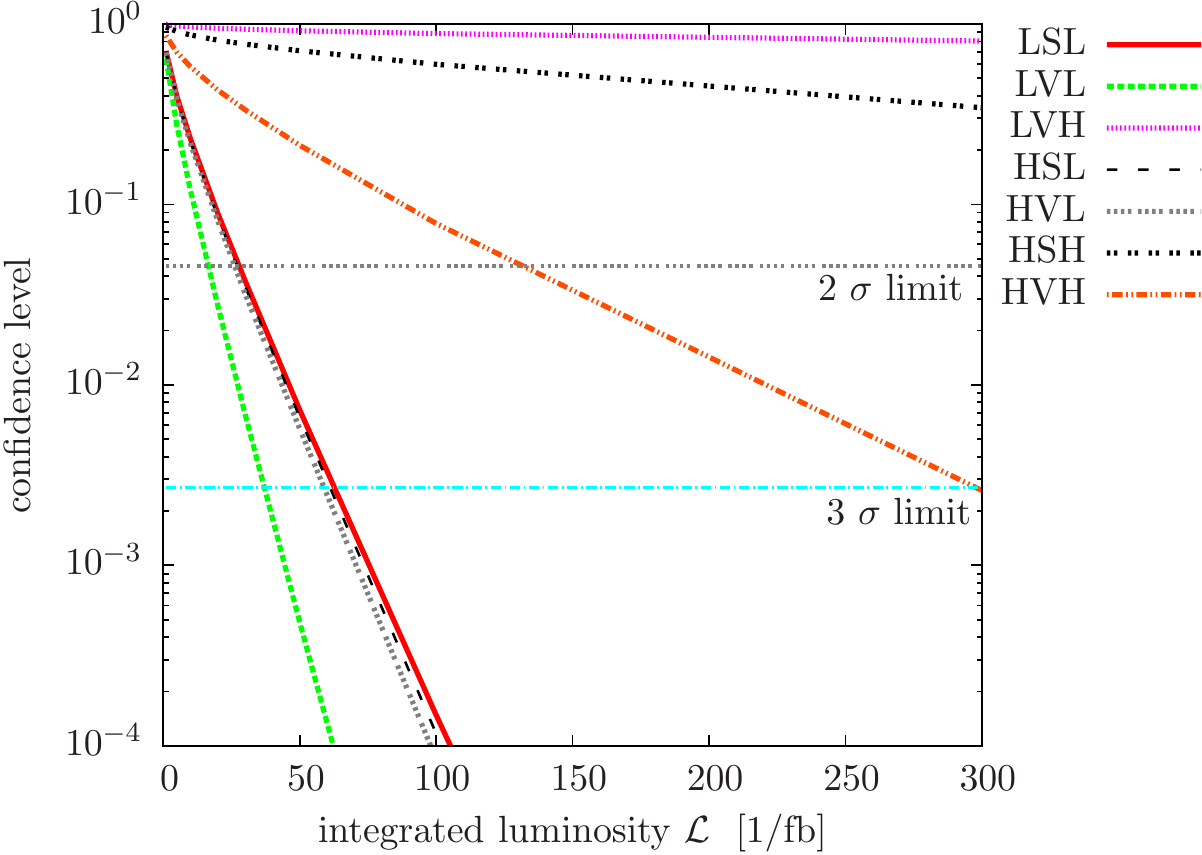} 
\label{sfigo25}
}
\hspace{0.2cm}
\subfigure[][~Invariant mass of $e^+e^-$ system]{
\includegraphics[width=0.43\textwidth]{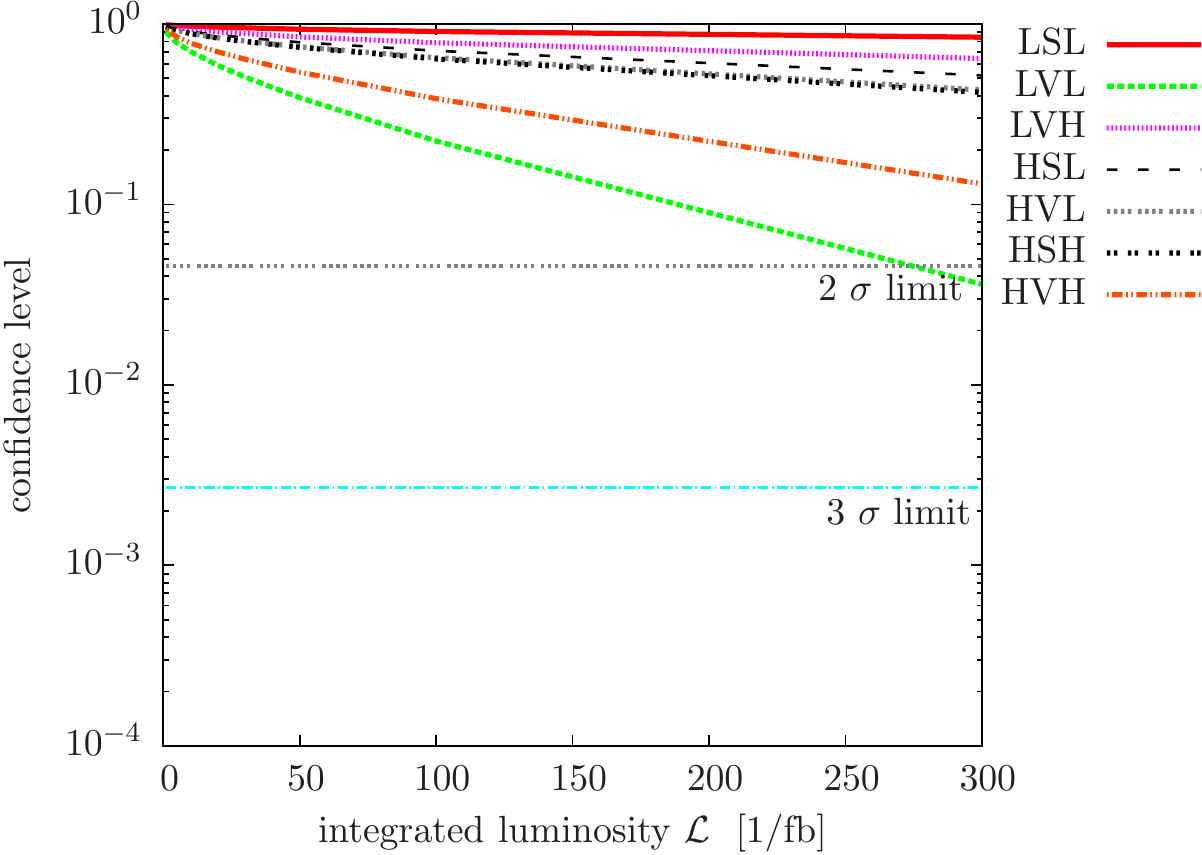} 
\label{sfigp25}
}
\caption{The curves illustrate the level to which the $M_{\mathrm{missing}}$ and $m_{e^+e^-}$ can discriminate between the different models, assuming the realization of LVH (\ref{sfigm25}-\ref{sfign25}), LSH (\ref{sfigo25}-\ref{sfigp25}). All cross sections are assumed to be $2.5 \%$ of the Standard Model background cross section.}

\label{fig:CLLSH} 
\end{figure*}

For the results shown in Figs.~\ref{fig:CLHSH}-\ref{fig:CLLSH} we assume respectively that one of the models is realized in nature, as indicated by the caption of each plot. The coupling strength in each model is chosen such that it is respecting present bounds. However, as discussed in Section~\ref{sec:eff}, bounds from direct and indirect detection experiments can be avoided in case the WIMPs are not stable on cosmological time scales. Only bounds from direct searches at LEP pose a stringent constraint. To evaluate how well the different quantum numbers of the benchmark models can be discriminated we assume all benchmark models have a cross section of $2.5\%$ of the Standard Model cross section. 

We have studied the impact of all the observables shown in Fig.~\ref{fig:CLSM}, namely missing transverse energy, $m_{e^+e^-}$, $\Delta \phi_{e^+e^-}$, $\Delta y_{e^+e^-}$, $p_T$ of the hardest lepton and $M_{\mathrm{missing}}$. We find that $m_{e^+e^-}$ and $M_{\mathrm{missing}}$ are sufficient to discriminate the quantum numbers of our candidate models confidently. $M_{\mathrm{missing}}$ gives a handle on the mass scales, while $m_{e^+e^-}$ discriminates the spin of the mediators. This is clearly demonstrated by comparing Figs.~\ref{sfigi25} and \ref{sfigj25}, where LVL and LSL are strongly discriminated by including $m_{e^+e^-}$ whereas $M_{\mathrm{missing}}$ has almost no discriminating power. The situation is reversed for the discrimination of LSL and LSH, as sown in Figs. \ref{sfigo25} and \ref{sfigp25}.  Here both, LSL and LSH have a pronounced peak in $m_{e^+e^-}$ at the Z boson mass (thus $m_{e^+e^-}$ is not discriminating the two models), but the large difference of the WIMP mass is reflected in $M_{\mathrm{missing}} \geq 2M_{\mathrm{WIMP}}$. However, if one wants to discriminate quantum numbers beyond mass and spin (e.g. CP structure) other observables should be included, e.g.  $\Delta \phi_{e^+e^-}$. A combination of the variables discussed will improve the statistical significance in discriminating the models' quantum numbers.

In general and in particularly for a leptophilic scenario, studying a dark sector using t-channel mediated forces is a challenging task at the LHC. This study has demonstrated that even with conservative assumptions on the level of polarization, the ILC can conclusively explore the quantum numbers of a dark sector by using a combination $M_{\mathrm{missing}}$  and $m_{e^+e^-}$. 

\begin{acknowledgments}
MR acknowledges partial support by the Deutsche
Forschungsgemeinschaft via the Sonderforschungsbereich/Transregio
SFB/TR-9 ``Computational Particle Physics'' and the Initiative and
Networking Fund of the Helmholtz Association, contract HA-101 (``Physics
at the Terascale'')
\end{acknowledgments}


\begin{thebibliography}{99}

\bibitem{Komatsu:2010fb}
  E.~Komatsu {\it et al.} [ WMAP Collaboration ],
  Astrophys.\ J.\ Suppl.\  {\bf 192}, 18 (2011).
  [arXiv:1001.4538 [astro-ph.CO]].

\bibitem{Picozza:2006nm} 
  P.~Picozza, A.~M.~Galper, G.~Castellini, O.~Adriani, F.~Altamura, M.~Ambriola, G.~C.~Barbarino and A.~Basili {\it et al.},
  Astropart.\ Phys.\  {\bf 27}, 296 (2007);
  O.~Adriani {\it et al.}  [PAMELA Collaboration],
  Nature {\bf 458}, 607 (2009).

\bibitem{FermiLAT:2011ab} 
  M.~Ackermann {\it et al.}  [Fermi LAT Collaboration],
  Phys.\ Rev.\ Lett.\  {\bf 108}, 011103 (2012)
  [arXiv:1109.0521 [astro-ph.HE]].

\bibitem{Aguilar:2013qda} 
  M.~Aguilar {\it et al.}  [AMS Collaboration],
  Phys.\ Rev.\ Lett.\  {\bf 110}, no. 14, 141102 (2013);
   Main results at http://www.ams02.org/.

\bibitem{Savage:2010tg} 
  C.~Savage, G.~Gelmini, P.~Gondolo and K.~Freese,
  Phys.\ Rev.\ D {\bf 83}, 055002 (2011);
 J.~M.~Cline, Z.~Liu and W.~Xue,
  Phys.\ Rev.\ D {\bf 87}, 015001 (2013).

\bibitem{Birkedal:2004xn}
  A.~Birkedal, K.~Matchev, M.~Perelstein,
  Phys.\ Rev.\  {\bf D70}, 077701 (2004).
  [hep-ph/0403004].
  
\bibitem{Feng:2005gj}
  J.~L.~Feng, S.~Su and F.~Takayama,
  Phys.\ Rev.\ Lett.\  {\bf 96}, 151802 (2006)
  [arXiv:hep-ph/0503117].

\bibitem{Beltran:2008xg}
  M.~Beltran, D.~Hooper, E.~W.~Kolb and Z.~C.~Krusberg,
  Phys.\ Rev.\  D {\bf 80}, 043509 (2009)
  [arXiv:0808.3384 [hep-ph]].

\bibitem{Cao:2009uw}
  Q.~-H.~Cao, C.~-R.~Chen, C.~S.~Li, H.~Zhang,
  JHEP {\bf 1108}, 018 (2011).
  [arXiv:0912.4511 [hep-ph]].
  
\bibitem{Beltran:2010ww}
  M.~Beltran, D.~Hooper, E.~W.~Kolb, Z.~A.~C.~Krusberg, T.~M.~P.~Tait,
  JHEP {\bf 1009}, 037 (2010).
  [arXiv:1002.4137 [hep-ph]].

\bibitem{Meade:2007js} 
  P.~Meade and M.~Reece,
  hep-ph/0703031.

\bibitem{Goodman:2010yf}
  J.~Goodman, M.~Ibe, A.~Rajaraman, W.~Shepherd, T.~M.~P.~Tait, H.~-B.~Yu,
  Phys.\ Lett.\  {\bf B695}, 185-188 (2011).
  [arXiv:1005.1286 [hep-ph]].

\bibitem{Bai:2010hh}
  Y.~Bai, P.~J.~Fox, R.~Harnik,
  JHEP {\bf 1012}, 048 (2010).
  [arXiv:1005.3797 [hep-ph]].

\bibitem{Rajaraman:2011wf}
  A.~Rajaraman, W.~Shepherd, T.~M.~P.~Tait, A.~M.~Wijangco,
  [arXiv:1108.1196 [hep-ph]].

\bibitem{Fox:2011fx} 
  P.~J.~Fox, R.~Harnik, J.~Kopp and Y.~Tsai,
  Phys.\ Rev.\ D {\bf 84}, 014028 (2011)
  [arXiv:1103.0240 [hep-ph]].

\bibitem{Fox:2011pm} 
  P.~J.~Fox, R.~Harnik, J.~Kopp and Y.~Tsai,
  Phys.\ Rev.\ D {\bf 85}, 056011 (2012).

\bibitem{Plehn:2001nj}
  T.~Plehn, D.~L.~Rainwater, D.~Zeppenfeld,
  Phys.\ Rev.\ Lett.\  {\bf 88}, 051801 (2002).
  [hep-ph/0105325].

\bibitem{Klamke:2007cu}
  G.~Klamke, D.~Zeppenfeld,
  JHEP {\bf 0704}, 052 (2007).
  [hep-ph/0703202 [HEP-PH]].

\bibitem{Andersen:2010zx}
  J.~R.~Andersen, K.~Arnold, D.~Zeppenfeld,
  JHEP {\bf 1006}, 091 (2010).
  [arXiv:1001.3822 [hep-ph]].

\bibitem{Eboli:2000ze}
  O.~J.~P.~Eboli, D.~Zeppenfeld,
  Phys.\ Lett.\  {\bf B495}, 147-154 (2000).
  [hep-ph/0009158].

\bibitem{Buras:2000dm}
  A.~J.~Buras, P.~Gambino, M.~Gorbahn, S.~Jager, L.~Silvestrini,
  Phys.\ Lett.\  {\bf B500}, 161-167 (2001).
  [hep-ph/0007085].

\bibitem{D'Ambrosio:2002ex}
  G.~D'Ambrosio, G.~F.~Giudice, G.~Isidori, A.~Strumia,
  Nucl.\ Phys.\  {\bf B645}, 155-187 (2002).
  [hep-ph/0207036].

\bibitem{Regge:1959mz}
  T.~Regge,
  Nuovo Cim.\  {\bf 14} (1959) 951.

\bibitem{Brower:1974yv}
  R.~C.~Brower, C.~E.~DeTar and J.~H.~Weis,
  Phys.\ Rept.\  {\bf 14} (1974) 257.

\bibitem{Aad:2008zzm}
  G.~Aad {\it et al.}  [ATLAS Collaboration],
  JINST {\bf 3} (2008) S08003.
  
\bibitem{Ball:2007zza}
  G.~L.~Bayatian {\it et al.}  [CMS Collaboration],
  J.\ Phys.\ G {\bf 34} (2007) 995.



\bibitem{Kaplan:2009ag} 
  D.~E.~Kaplan, M.~A.~Luty and K.~M.~Zurek,
  Phys.\ Rev.\ D {\bf 79}, 115016 (2009)
  [arXiv:0901.4117 [hep-ph]].

\bibitem{Fox:2008kb} 
  P.~J.~Fox and E.~Poppitz,
  Phys.\ Rev.\ D {\bf 79}, 083528 (2009)
  [arXiv:0811.0399 [hep-ph]].

\bibitem{Cirelli:2008pk} 
  M.~Cirelli, M.~Kadastik, M.~Raidal and A.~Strumia,
  Nucl.\ Phys.\ B {\bf 813}, 1 (2009)
  [Addendum-ibid.\ B {\bf 873}, 530 (2013)]
  [arXiv:0809.2409 [hep-ph]].

\bibitem{Chen:2008dh} 
  C.~-R.~Chen and F.~Takahashi,
  JCAP {\bf 0902}, 004 (2009)
  [arXiv:0810.4110 [hep-ph]].

\bibitem{An:2013xka} 
  H.~An, L.~-T.~Wang and H.~Zhang,
  arXiv:1308.0592 [hep-ph].

\bibitem{Cohen:2009fz} 
  T.~Cohen and K.~M.~Zurek,
  Phys.\ Rev.\ Lett.\  {\bf 104}, 101301 (2010)
  [arXiv:0909.2035 [hep-ph]].

\bibitem{Strassler:2006im} 
  M.~J.~Strassler and K.~M.~Zurek,
  Phys.\ Lett.\ B {\bf 651}, 374 (2007)
  [hep-ph/0604261].

\bibitem{Dreiner:2012xm} 
  H.~Dreiner, M.~Huck, M.~Kramer, D.~Schmeier and J.~Tattersall,
  Phys.\ Rev.\ D {\bf 87}, 075015 (2013)
  [arXiv:1211.2254 [hep-ph]].

\bibitem{ILCTDR}
  ILC Technical Design Report [ILC Collaboration], http://www.linearcollider.org/ILC/Publications/Techn\linebreak ical-Design-Report; 
  G.~Aarons {\it et al.}  [ILC Collaboration],
  arXiv:0709.1893 [hep-ph].

\bibitem{Abel:2008ai} 
  B.~Holdom,
  Phys.\ Lett.\ B {\bf 166}, 196 (1986);
   R.~Foot and X.~-G.~He,
  Phys.\ Lett.\ B {\bf 267}, 509 (1991);
  S.~A.~Abel, M.~D.~Goodsell, J.~Jaeckel, V.~V.~Khoze and A.~Ringwald,
  JHEP {\bf 0807}, 124 (2008);
   N.~Arkani-Hamed and N.~Weiner,
  JHEP {\bf 0812}, 104 (2008).
  


\bibitem{llhr}
  T.~Junk,
  Nucl.\ Instrum.\ Meth.\  A {\bf 434} (1999) 435.
  T.~Junk,
  CDF Note 8128 [cdf/doc/statistics/public/8128].
  T.~Junk,
  CDF Note 7904 [cdf/doc/statistics/public/7904].
  H. Hu and J. Nielsen, 
  in 1st Workshop on Confidence Limits', 
  CERN 2000-005 (2000). 


\bibitem{Read:2002hq}
  A.~L.~Read,
  CERN-OPEN-2000-205.  
  A.~L.~Read,
  J.\ Phys.\ G {\bf G28 } (2002)  2693-2704.


\end{thebibliography}
\end{document}